TOPICAL REVIEW

# Specialized machine learning force fields for materials dynamics

**Yue Wu[1,2], Haoyu Wan[1], Yuan Tian[2], Deng Pan[1,*] and Mingwei Chen[2,*]**

[1] Materials Genome Institute, Shanghai University, Shanghai 200444, China

[2] Department of Materials Science and Engineering, Southern University of Science and Technology, Shenzhen 518055, China

[*] Authors to whom any correspondence should be addressed.

E-mail: DPan_MGI@shu.edu.cn (Deng Pan); chenmw@sustech.edu.cn (Mingwei Chen)

Author e-mails: repeatsongyue@gmail.com (Yue Wu); wanhaoyu0715@gmail.com (Haoyu Wan); tiany3@sustech.edu.cn (Yuan Tian)

### Abstract

Machine learning interatomic potentials (MLIPs) are transforming the length and time scales accessible to atomistic simulation. Equivariant graph neural networks, combined with large first-principles datasets, have produced pretrained models with broad chemical coverage and zero-shot performance for static and near-equilibrium properties. Translating this breadth into reliable performance in complex dynamics, however, requires validation for the intended conditions and observables. This Review examines how physics-driven specialization extends the reach of MLIPs in complex materials dynamics. Three considerations guide this assessment. First, training distributions dominated by near-equilibrium structures can leave reactive or highly strained configurations undersampled, while representation bias and electronic-structure reference errors impose distinct limitations. Second, the computational cost of high-order equivariant architectures and multi-hop message passing can restrict accessible system sizes and simulation times. Third, field-coupled and open-system applications may require additional state variables or nonlocal interactions; force conservativity is a separate architectural choice, independent of chemical coverage. Through four overlapping application contexts—electrified interfaces under constant potential, open systems with fluctuating particle numbers, phase transformations and multiphase evolution, and large-scale deformation and fracture—we connect these requirements to observable-specific validation. Foundation models and specialized potentials occupy complementary positions in this design space. A pretrained model can be used directly where it meets the target requirements; the benefits of targeted adaptation should be established through the intended observables and suitable task-compatible baselines. We conclude with a roadmap for physically consistent, hardware-aware force fields that connect electronic-structure references to validated atomistic mechanisms and larger-scale materials models.

## 1. Introduction

In modern computational materials science and chemical engineering, a fundamental understanding of atomic-scale microstructural evolution is the cornerstone for elucidating macroscopic properties, such as catalytic dynamics, phase transition mechanisms, and mechanical failure. Since the formulation of the Born-Oppenheimer approximation, the accurate and efficient construction of the potential energy surface (PES) has remained a central and demanding task in molecular simulations.

On the one hand, first-principles methods, predominantly density functional theory (DFT) and ab initio molecular dynamics (AIMD), describe atomic interactions through electronic-structure calculations; in standard Born–Oppenheimer AIMD, the electrons follow the nuclei adiabatically rather than undergoing explicit nonadiabatic dynamics [1, 2]. However, the computational complexity of standard Kohn-Sham DFT typically scales cubically with the number of electrons ( $\mathcal{O}(N^3)$ ). Although linear-scaling ( $\mathcal{O}(N)$ ) methods based on approximations such as localized Wannier functions have been developed, their efficiency depends on electronic locality, matrix decay, numerical tolerances, and the system size at which reduced asymptotic scaling offsets implementation overhead [3]. This stringent scaling relation imposes a formidable computational bottleneck, often restricting routine AIMD simulations to spatial domains of hundreds of atoms and temporal scales of tens of picoseconds (ps). Consequently, direct first-principles dynamical simulations remain computationally prohibitive for rare events involving high-energy barrier crossings, the prolonged relaxation of complex solid-liquid interfaces, and the slow kinetic processes inherent to heterogeneous catalysis [4].

On the other hand, to overcome these spatiotemporal limitations, the field has long sought to develop empirical interatomic potentials (EIPs), aiming to reduce computational complexity to linear scaling ($\mathcal{O}(N)$) via analytic functional forms. This trajectory represents a progressive evolution from simple to complex formulations. Early models predominantly relied on pair potentials, such as the Lennard-Jones potential for dispersion and short-range repulsion [5], and ionic pair-potential forms that combine long-range Coulomb interactions with Born–Mayer or Buckingham short-range repulsion [6, 7]. However, pure pair potentials generally cannot represent the directional bonding and environment-dependent energetics required for many real crystals and defects. Consequently, researchers introduced many-body terms incorporating angular and directional dependencies. This led to the development of the Stillinger-Weber (SW) potential [8] and the bond-order-based Tersoff potential [9] for covalent semiconductor systems, as well as the embedded-atom method (EAM) [10], specifically designed to approximate the embedding of atoms in the electron density of metals.

To further simulate the critical processes of chemical bond dissociation and formation within a classical framework, researchers extended the bond-order concept to develop reactive force fields, such as ReaxFF [11], and the Charge-Optimized Many-Body (COMB) potential, which incorporates dynamic charge equilibration and electronegativity equalization mechanisms [12, 13].

Empirical interatomic potentials (EIPs), including reactive force fields, rely on chosen analytical forms and fitted parameters. These choices can limit flexibility and transferability outside the chemical environments and thermodynamic conditions used for parameterization, although a well-designed empirical model can remain highly effective within its domain. Reactive PESs contain many minima and saddle regions, so extending one fixed functional form across diverse chemistries can require substantial expert effort and validation [14]. MLIPs offer a more flexible alternative, but they exchange explicit functional-form assumptions for assumptions about representation, data, optimization, and extrapolation.

Confronted with the dual challenges of limited transferability and analytical complexity in physical modeling, the field has undergone a data-driven paradigm shift. Machine learning interatomic potentials (MLIPs), acting as surrogate models for first-principles calculations, replace many hand-parameterized interaction terms with flexible learned functions, while still requiring architectural choices such as descriptors, symmetry constraints, cutoffs, and energy decompositions. Universal-approximation results motivate the expressive capacity of neural networks and kernel methods under appropriate assumptions. Data efficiency, extrapolation reliability, and physical consistency depend on the representation and training procedure. MLIP construction is therefore a constrained regression problem: learning mappings from atomic configurations to energies and related derivatives while respecting the relevant symmetries and conservation laws [15].

A key step was to represent the many-atom potential energy surface through local atomic environments. In 2007, Behler and Parrinello introduced the landmark high-dimensional neural network potential (HDNNP) architecture [16]. This framework incorporated the key assumption of "local energy decomposition," which posits that the total energy of a system can be approximated as the linear sum of energy contributions from individual local atomic environments ($E = \sum_{i=1}^{N} E_i\,(\mathcal{R}_i)$). By mapping the topological neighborhood of each atom into translationally, rotationally, and permutationally invariant symmetry functions (e.g., ACSF), HDNNP converts PES fitting into local-environment regression problems that can be evaluated in parallel, subject to the validity of the locality approximation for the target system. Subsequently, the Gaussian approximation potential (GAP) proposed by Bartók et al., combined with the SOAP descriptor, further established a rigorous mathematical formalism for this local mapping within a kernel regression framework [17].

Building upon this local-energy representation, MLIPs have achieved a long-sought dual breakthrough in the physics community. In terms of accuracy, within a sufficiently sampled domain, MLIPs can reproduce the PES of a chosen electronic-structure reference (e.g., DFT or, for suitably small training systems, CCSD(T)) with small interpolation errors. In terms of efficiency, they bypass self-consistent electronic-structure iterations. Local models with fixed cutoffs and bounded neighbor counts can achieve asymptotically linear scaling with atom number, whereas models with explicit long-range interactions, global attention, or charge equilibration may have different scaling and communication costs [18].

This characteristic endows MLIPs with substantial capabilities. Large-scale complex simulations previously precluded in AIMD due to computational constraints, such as fracture dynamics in materials comprising millions of atoms, microsecond-scale solid-liquid phase transitions, and in situ reaction network exploration on heterogeneous catalytic surfaces, can now be realized on supercomputers or even single-node GPUs [14, 19]. Consequently, MLIPs help bridge the spatiotemporal gap between "microscopic electronic structures" and "mesoscopic thermodynamic statistics" in multiscale physical modeling.

It is important to acknowledge that recent optimizations in underlying network operators and the accumulation of massive datasets have successfully catalyzed the development of "universal machine learning interatomic potentials" (universal MLIPs) that span the periodic table (e.g., M3GNet [20], CHGNet [21], and MACE-MP-0 [22]). Supported by high-throughput materials databases (such as the Materials Project or OQMD) [23, 24, 25], these foundation models have demonstrated useful zero-shot generalization capabilities in ground-state (0 K) static structural relaxations, lattice elastic property predictions, and near-equilibrium phonon spectra calculations, thereby substantially accelerating the throughput of static materials discovery.

However, when the research focus shifts from static materials property screening to complex chemical dynamics involving drastic configurational evolution, the reliance on static ground-state data reveals underlying physical limitations. Realistic complex chemical evolutions, such as ion desolvation at battery interfaces under constant operating voltages, composition-changing surface growth and reconstruction, or bond dissociation in heterogeneous catalysis, often involve abrupt changes in atomic coordination and require the system to traverse high-energy transition states (TS) [26]. For models trained predominantly on low-energy structures and relaxation trajectories, this data-distribution "cold bias" can limit reliability when simulations visit poorly sampled reactive or highly strained configurations [27].

In these domains, equivariance or a larger receptive field does not by itself establish reliability beyond the sampled configurations. In reactive molecular dynamics (MD), extrapolation beyond the sampled configurations can produce an unphysical PES and unstable trajectories [28]. In essence, the performance ceiling of modern MLIPs in reactive systems is dictated not only by the expressive capacity of the network architecture but also fundamentally constrained by the physical sampling fidelity of the training datasets. For reactive applications, sampling relevant pathways and high-energy environments can improve coverage; the required configurations depend on the target process [26].

Reliable dynamics requires separate checks of extrapolation error and energy drift. Local-energy models should preserve extensivity, with forces derived from the modeled energy. NVE drift also depends on PES smoothness, integrator, timestep, constraints, and numerical precision [29, 30]. Long-range electrostatics and polarization require separate treatment when relevant. Broader sampling cannot correct systematic electronic-structure reference errors.

This review connects model design and sampling to the physical requirements of materials dynamics. Foundation models provide broad interaction priors and can serve directly as potentials in validated domains [22, 31]. The focus is on how targeted sampling, physical variables, and efficient deployment support physics-driven specialization for demanding applications. Figure 1 organizes the available design choices, including symmetry-aware representations, targeted sampling, explicit ensemble variables, and physical validation. The subsequent sections review architectural development (Section 2), targeted sampling (Section 3), model selection and adaptation in four application contexts (Section 4), and scientific-consistency criteria (Section 5). Together, these sections link application requirements to observable-specific tests and deployment constraints.

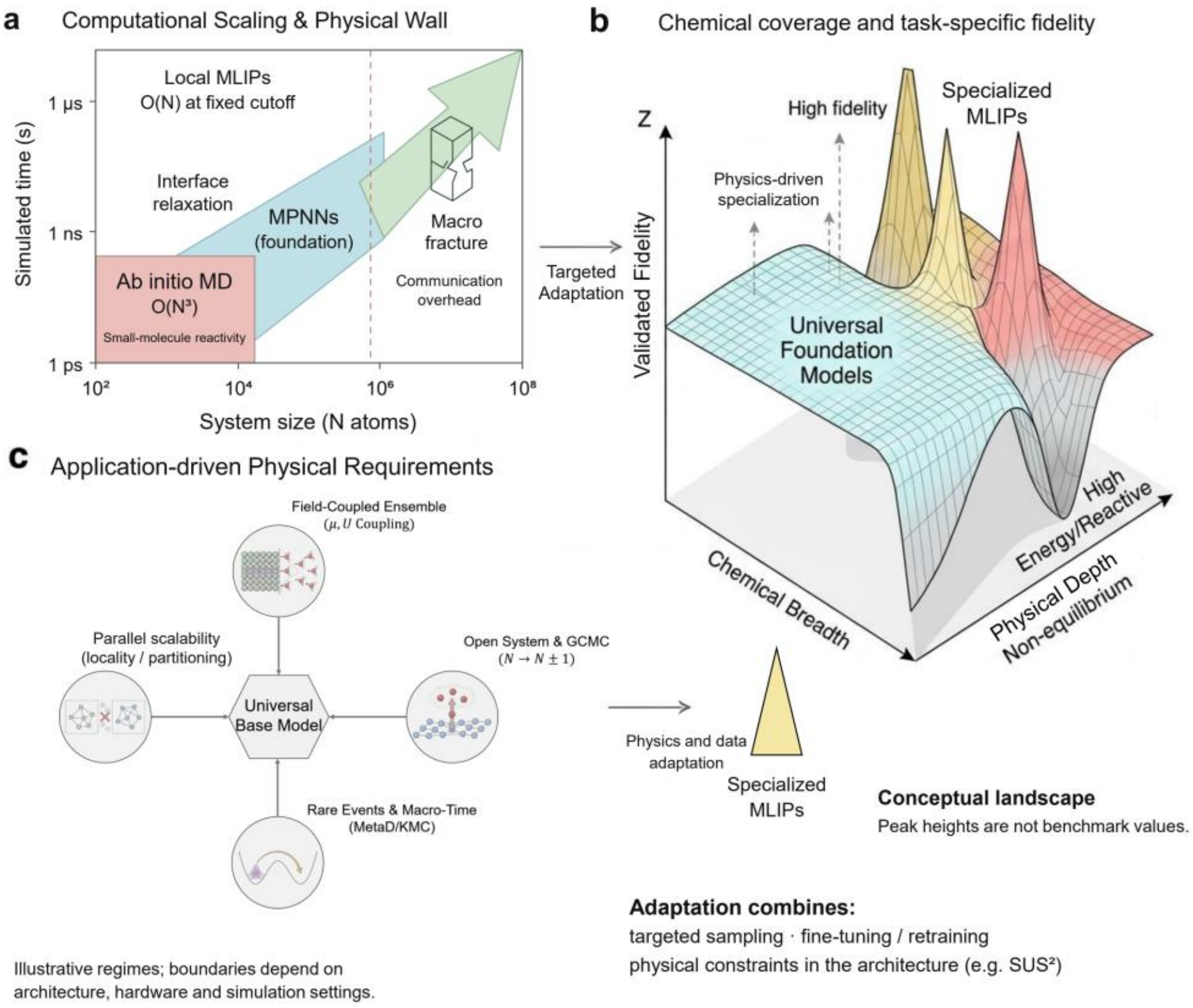

Figure 1. Physics-driven specialization for materials dynamics. (a) Spatiotemporal scaling limits of atomistic simulations. Conventional Kohn–Sham ab initio molecular dynamics (MD) commonly exhibits cubic scaling in system size, while local machine-learning potentials expand the accessible simulation envelope. Deep message-passing models can incur substantial distributed-memory communication overhead, whereas shallow or strictly local architectures can improve parallel scalability and enable much larger simulations (e.g., fracture). (b) Conceptual phase diagram illustrating the capability landscape of MLIPs. Broad chemical coverage (X-axis) and coverage of demanding physical conditions (Y-axis) are distinct from validated predictive fidelity (Z-axis). The peaks schematically indicate possible gains in task-specific fidelity from targeted adaptation. (c) Adaptation route for applications requiring additional physical or configurational coverage. Depending on the target observable, model construction and simulation may need to address: external field coupling (for $\mu, U$ electrochemical interfaces), open-system GCMC (permitting particle fluctuations $N \rightarrow N \pm 1$), macro-time acceleration (MetaD/KMC for rare events), and locality-aware or communication-aware parallel evaluation (to reduce distributed-memory overhead). Adaptation may combine targeted sampling, fine-tuning or retraining, and physical constraints in the architecture. The landscape and accessible simulation regimes are schematic rather than quantitative benchmarks.

## 2. Representations and network architectures

To overcome the generalization bottlenecks of empirical physical modeling, MLIPs have undergone a rapid revolution in their underlying mathematics and network architectures over the past decade. These complementary approaches span “local scalars → topological networks → 3D equivariant tensors → multidimensional physical degrees of freedom.” Their relative accuracy and cost depend on the target system, training data, and implementation (Figure 2).

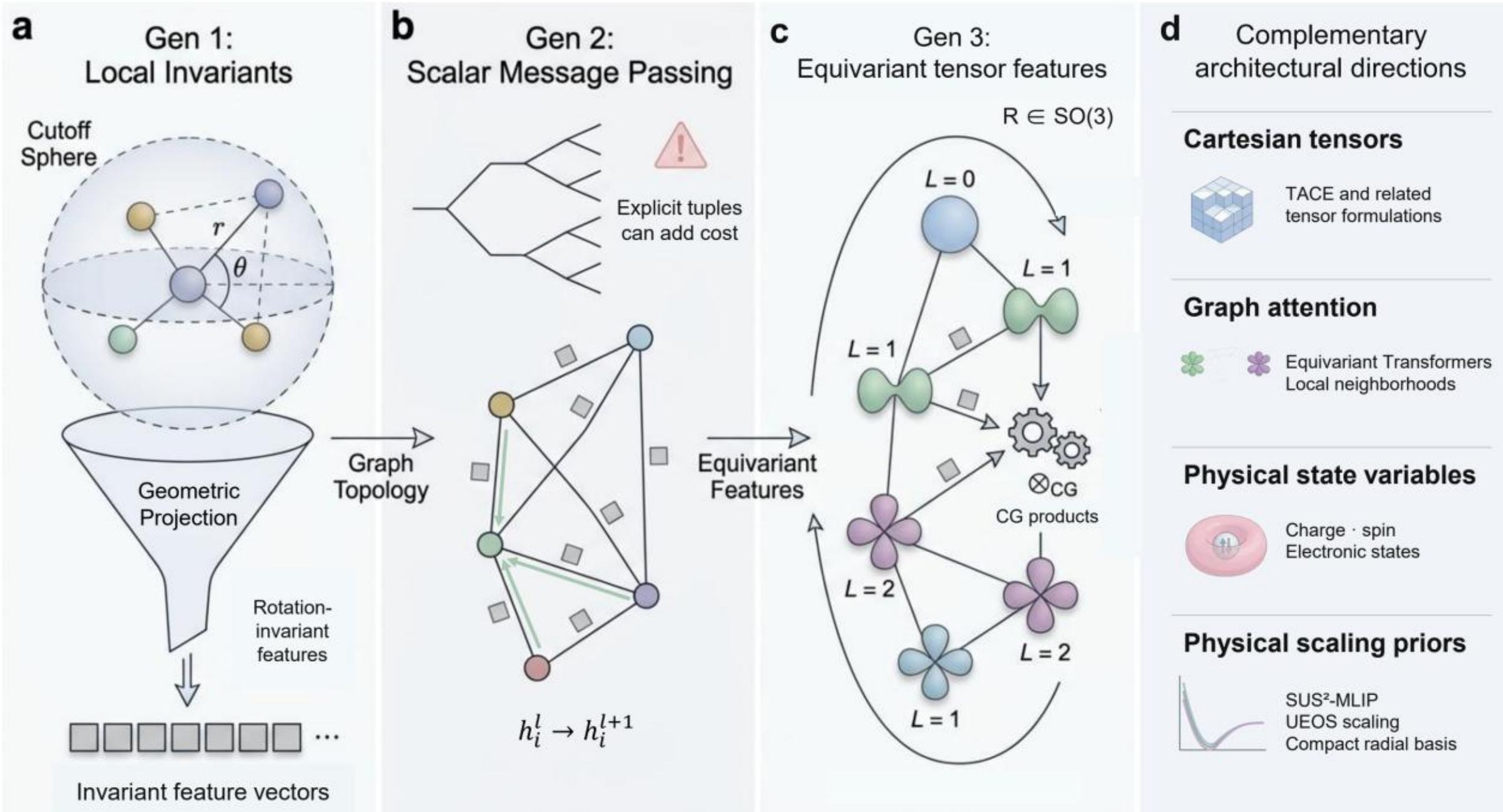

Figure 2. Selected historical developments in mathematical representations and network architectures in MLIPs. (a) First-generation local invariant descriptors (e.g., ACSF, SOAP). Three-dimensional geometric information (distances $r$, angles $\theta$) within a local cutoff sphere is encoded in scalar invariant descriptors. These remove dependence on global orientation while retaining information about internal geometry; their ability to distinguish environments depends on construction and truncation. (b) Second-generation scalar message-passing networks. Atomic graphs enable information flow, while explicit enumeration of angles or torsions can add costs that scale with the number of neighbor pairs or triplets; the precise complexity depends on coordination, cutoff, and implementation. (c) Third-generation equivariant graph neural networks (eGNNs) based on spherical tensors. Node embeddings are expanded into irreducible representations of $SO(3)$ (for example, channels with $l = 0,1,2$). Under a rotation of the coordinates, tensor features transform according to the corresponding representations. Many such models use computationally demanding Clebsch–Gordan tensor products ($\otimes_{CG}$), although this is not the only construction of equivariance. (d) Selected frontier directions. Cartesian-tensor formulations such as TACE provide alternatives to spherical Clebsch–Gordan implementations; equivariant Transformers improve attention-based interaction modeling but are often still cutoff-local; and electronic-state, charge, or spin variables extend models beyond a single geometry-only adiabatic PES. Physical scaling priors provide a complementary architectural constraint, illustrated here by universal-equation-of-state (UEOS) scaling in the radial representation of SUS²-MLIP.

## 2.1. Local invariant representations

When introducing machine learning to atomic-scale simulations, the central mathematical challenge is that raw Cartesian coordinates depend on the reference frame, whereas the model must obey the relevant physical symmetries. This is because the total energy and atomic forces of a system must strictly preserve invariance (or equivariance) under translations and rotations of the reference frame (i.e., $E(3)$ symmetry), and the permutation of identical atoms should not alter the macroscopic state. Consequently, the first generation of MLIPs established a two-step paradigm: "geometric feature engineering followed by nonlinear regression." Depending on the choice of regressor (neural networks vs. kernel methods) and the local environment mapping strategy, this generation diverged into two complementary evolutionary branches.

The first is Atom-Centered Symmetry Functions (ACSF) and High-Dimensional Neural Networks. The foundational work for this branch was proposed by Behler and Parrinello in 2007 [16]. They introduced the assumption of "local energy decomposition," which approximates the total energy of a macroscopic system as the scalar sum of energy contributions from individual local atomic environments. To enable neural networks to "comprehend" these local environments, they designed ACSF. Using radial and angular symmetry functions, ACSFs map the local three-dimensional geometry to a fixed-length vector of continuous, rotationally and translationally invariant values (Figure 2a).

HDNNPs based on finite-cutoff ACSFs have asymptotically linear inference cost in system size when the atomic density and number of neighbors per atom remain bounded. Their actual cost also depends on neighbor-list construction, species count, and implementation. These architectures provided important methodological foundations for later frameworks such as Deep Potential, which introduced learned local-environment representations.

Bartók et al. proposed the Gaussian Approximation Potential (GAP) in 2010 [17], subsequently introducing the Smooth Overlap of Atomic Positions (SOAP) descriptor [32]. SOAP expands a smooth neighbor density in radial and spherical-harmonic basis functions. By constructing a power spectrum, it naturally integrates out the rotational degrees of freedom to achieve rotational invariance.

As a representative non-parametric kernel method, the GAP/SOAP framework established an accuracy benchmark in materials science [33, 34]. Its core advantages include data efficiency and a model-based uncertainty estimate. In data-scarce regimes, kernel methods can approximate reference electronic-structure targets using hundreds to thousands of configurations; for example, SOAP-based Gaussian process regression has been used to predict hydrogen-atom-transfer barriers in proteins [35]. Gaussian-process posterior variances can support active learning, but they are calibrated uncertainty measures only to the extent that the kernel, noise model, and training distribution are appropriate. Jinnouchi et al. integrated Bayesian error indicators into VASP for on-the-fly training [36], and Caro et al. used Gaussian-process uncertainty estimates to assess local environments encountered during amorphous-carbon deposition [37]. These applications demonstrated the robustness of GAP in several complex solid-state settings (e.g., defect evolution in tungsten [38], polymorphic silicon systems [39], and amorphous carbon networks [40]). This robustness reflects both descriptor expressivity and targeted sampling, providing an early paradigm for the physics-targeted strategies discussed in Section 3.

Related whole-molecule kernel approaches include gradient-domain machine learning (GDML), which learns conservative molecular force fields from reference forces [41]. Its symmetry-adapted extension, sGDML, incorporates molecular permutation symmetries and was demonstrated using coupled-cluster reference data for selected molecules [42].

After ACSF and SOAP established the paradigm of local invariant features, the field faced an engineering trade-off: neural networks can require substantial data and careful optimization, whereas full Gaussian-process training scales cubically with the number of training environments unless sparse approximations are used. GAP inference in molecular dynamics instead scales with the number of retained sparse points and local environments. To balance expressive capacity and computational efficiency, force-field frameworks based on high-order features combined with linear regression emerged. An early representative of this branch is the Spectral Neighbor Analysis Potential (SNAP) proposed by Thompson et al. in 2015 [43]. The core concept of SNAP involves projecting local atomic densities onto a 4D hypersphere and expanding them via 4D hyperspherical harmonics to construct bispectrum components as rotationally invariant descriptors. Requiring only linear parameter fitting, SNAP demonstrated high execution speeds. Wood and Thompson extended SNAP with quadratic combinations of bispectrum components, finding that larger training sets were needed to control overfitting [44]. An explicit multielement formulation subsequently improved chemical resolution [45]. The W–Be helium-bubble application is discussed with its original study in Section 4.3.

Moment tensor potentials (MTPs) construct invariant basis functions by contracting local moment tensors [46]. D-optimal active learning uses the linear design matrix to identify extrapolative configurations [47]; its extrapolation grade is a model-dependent selection criterion. The MLIP package supports this workflow [48]. Applications include multicomponent-alloy searches [49], gas-phase reactions such as S + $H_2$ [50], and promoted Ag catalysts for ethylene epoxidation [51].

The atomic cluster expansion (ACE) [52] and related polynomial invariants [53] provide systematic bases for local atomic energies. Completeness results apply under specified regularity, cutoff, and truncation conditions [54]. Efficient recursive evaluation, as implemented in PACE, reduces the practical cost of higher-order bases [55]. Applications span organic conformations [56], $NH_3$ dissociation [57], and Diels–Alder reactions in solution [58]. Their accuracy and efficiency depend on basis size and the sampled domain, making compact expansions useful for selected reactive tasks.

Alongside linear and polynomial expansions such as SNAP and ACE, neural networks provided flexible nonlinear mappings from local descriptors to atomic energies.

As early as the 1990s, Blank et al. attempted to directly map atomic coordinates to the PES using neural networks [59]. These early fits used restricted sets of degrees of freedom, including CO/Ni(111) and a 12-dimensional $H_2$/Si(100) interaction surface, establishing neural networks as potential-energy surrogate models. A key development came in 2007 when Behler and Parrinello’s BPNN (the aforementioned HDNNP) established the modern paradigm of "local descriptors + element-specific neural networks" [16]. In 2011, Behler systematically derived the mathematical forms of various radial and angular symmetry functions [60], which laid the foundation for neural network force fields over the subsequent decade. BPNNs have found broad application at complex interfaces; for example, Romano et al. used BPNNs to simulate interfacial water dissociation, proton transfer, and water-layer organization at the magnetite-liquid water interface [61]. For ionic systems, Ghasemi et al. combined neural-network predictions of environment-dependent atomic electronegativities with global charge equilibration to determine the charge distribution and energy [62].

However, traditional BPNNs relied heavily on tedious parameter tuning tailored to specific material systems. To propel neural network force fields toward true "general-purpose" applicability, Smith et al. proposed the ANI-1 potential (ANAKIN-ME) in 2017 [63]. By modifying ACSF to capture richer angular information, ANI successfully achieved generalization across vast organic molecular spaces. Subsequently, this series achieved improvements in predictive accuracy and chemical coverage: Smith et al. introduced transfer learning, using inexpensive DFT data for pre-training to successfully fine-tune the general neural network (ANI-1ccx) to the highly expensive CCSD(T)/CBS coupled-cluster accuracy standard [64]; Devereux et al. then extended the architecture to complex systems containing sulfur and halogens (S, F, Cl) with ANI-2x [65]. Coupled with the efficient, open-source automatic differentiation platform TorchANI [66], non-GNN architectures based on refined descriptors provided practical tools for molecular modeling and conformational searches. A related molecular modeling approach, AIMNet2, combines iterative message passing, charge updates constrained by the total molecular charge, and explicit electrostatic and dispersion terms to describe neutral and charged species [67].

For materials dynamics, the neuroevolution potential (NEP) offers another efficient local neural-network formulation. NEP combines radial and angular invariant descriptors with a neural network trained using a separable natural evolution strategy [68]. Its GPU-oriented implementation in GPUMD integrates potential construction with atomistic simulation, including heat-transport calculations [69]. This approach complements message-passing architectures when local descriptors meet the required accuracy and high simulation throughput is a priority.

Deep Potential (DP) introduced learned representations of local atomic environments [70]. DeepPot-SE provided a smooth embedding suitable for dynamical simulations [71], and DeePMD-kit enabled training and deployment [72]. Standard DeepPot-SE remains a finite-cutoff local model unless supplemented with nonlocal interactions. Its large-scale implementations are discussed separately in Section 4.5.

Standard short-range forms of ACSF, SOAP, ACE, and DeepPot-SE share a finite spatial cutoff. This supports efficient local evaluation at bounded density. It also limits the information available to describe interactions whose energetics depend on the wider environment. The consequences differ for chemical reactivity, long-range response, and tensorial observables.

Long-range interactions also limit the chemical context available to a local model. In non-uniform systems (e.g., polarized interfaces, charged defects, heterogeneous catalytic surfaces), Coulombic electrostatic forces and van der Waals dispersion forces decay very slowly. Strict local truncation gives the model limited receptive field, rendering it unable to perceive long-range fluctuations in charge density beyond the cutoff sphere. Consequently, predictions of dielectric responses and macroscopic polarizability are prone to failure.

Local descriptors do not imply fixed chemical bonds. Neighbor lists and descriptors can change continuously as atoms move, allowing local MLIPs to describe bond breaking and formation. The remaining limitation concerns information beyond the receptive field and configurations outside the validated training domain. Message passing can broaden that field, but is not a prerequisite for chemical reactivity.

Tensorial responses require an appropriate output construction. Scalar invariant features can contribute through symmetry-adapted readouts or derivatives with respect to external fields. Equivariant features provide a complementary route that represents transformation rules within the network.

Zuo et al. compared GAP, MTP, neural-network potentials, linear SNAP, and quadratic SNAP across six elemental systems, illustrating model- and dataset-dependent accuracy–cost trade-offs [73].

First-generation MLIPs remain effective for many system-specific problems but often require deliberate sampling and parameterization. Finite-cutoff descriptors without explicit nonlocal variables cannot represent interactions that depend on global charge redistribution or polarization. Message-passing GNNs expand the effective receptive field and reduce some manual feature engineering. Long-range electrostatics and extrapolation remain separate modeling challenges.

## 2.2. Scalar message-passing networks

The "localized information" and "fixed cutoff" bottlenecks exposed by first- and quasi-second-generation potentials when handling complex topological reorganizations and long-range physical responses stem from the strong spatial locality assumption inherent in their local descriptors (whether hand-crafted or adaptively embedded). To extend the spatial context available to each atomic representation, the computational materials community introduced the Graph Neural Network (GNN) paradigm, marking the formal entry of MLIPs into the second generation of "end-to-end" dynamic feature learning.

An atomic configuration can be represented as a graph $\mathcal{G} = (\mathcal{V}, \mathcal{E})$, with nodes $\mathcal{V}$ encoding atomic species and edges $\mathcal{E}$ connecting selected neighbors. Message-passing neural networks (MPNNs) iteratively aggregate and update these features [74]. The connectivity is generally constructed from atomic positions rather than a fixed list of chemical bonds.

Each iteration of message passing allows an atom to acquire information from a broader graph neighborhood. Kang et al. showed that extrapolation behavior depends strongly on the mathematical form of the message-passing update [75]. Stacking $L$ local layers extends the receptive field to at most $L$ graph hops, with interactions beyond that range requiring additional treatment.

Early GNNs predominantly relied on two-body distance scalars as edge features to naturally guarantee global rotational and translational invariance. SchNet, proposed by Schütt et al. in 2017, serves as the foundational work in this direction [76]. SchNet discarded pre-defined symmetry functions, utilizing continuous-filter convolutional neural networks (CNNs) to directly learn many-body interactions within the distance space, achieving high computational throughput.

Scalar networks can be supplemented with additional physical terms and variables. PhysNet includes electrostatics and dispersion corrections [77, 78] and has been used for Diels–Alder dynamics [79]. SchNarc combines SchNet with surface hopping for excited-state photochemistry [80]. FieldSchNet includes external-field information and has modeled solvent effects on a Claisen rearrangement barrier [81]. These applications extend the modeled physics while retaining task-specific requirements for training and validation.

To endow scalar networks with directional awareness, Klicpera et al. proposed DimeNet, shifting the message-passing target from nodes to directed edges and using spherical Bessel functions and spherical harmonics to encode interatomic distances and bond angles [82, 83].

Subsequently, SphereNet [84] further incorporated higher-order geometric information, including torsional angles.

M3GNet [20] and CHGNet [21] extended scalar graph models to broad materials datasets. CHGNet used local magnetic moments as auxiliary training information to improve its representation of chemical environments. These models support structural relaxation and property screening; their dynamical validity depends on the process. The Direct Integration of the External Potential (DIEP) representation supplied another physics-informed embedding, with applications to fracture onset in carbon nanotubes [85]. Such examples motivate mechanism-specific evaluation alongside general benchmarks.

The geometric extensions in DimeNet/SphereNet and the broader chemical coverage of M3GNet/CHGNet illustrate the capabilities of scalar graph networks. Their design involves trade-offs in geometric resolution, representation, and evaluation cost:

Explicit angle and dihedral tuples can increase evaluation cost with coordination number. The scaling depends on tuple construction, cutoff, and implementation. Dense environments may therefore make higher-order scalar graph models expensive, even when their geometric resolution improves predictions (Figure 2b).

Scalar hidden states do not carry explicit non-scalar irreducible representations. A scalar energy model can still produce rotationally equivariant forces by differentiation, and those forces are conservative when derived from a differentiable invariant energy. Generic vector or higher-rank tensor observables require an appropriate symmetry-aware output construction or equivariant features.

These costs motivated architectures whose internal features transform equivariantly, allowing directional information to be propagated without explicit enumeration of every high-order tuple. Equivariant GNNs based on spherical tensors and group representation theory are a prominent solution, while scalar, Cartesian, and hybrid constructions remain viable depending on the task.

### 2.3. Spherical-tensor equivariant networks

Equivariant graph neural networks (eGNNs) provide a different inductive bias for representing atomic environments. Invariant descriptors can encode internal angles and anisotropic arrangements, and differentiating a smooth invariant energy yields equivariant forces. Equivariant networks additionally propagate vector and higher-order tensor features with prescribed rotational transformation rules. This changes how geometric information is represented and exchanged and can improve data efficiency, without making non-scalar hidden features a prerequisite for physically consistent energies and forces.

Under rotation, energy remains invariant while forces and tensorial responses transform with the coordinate frame. Spherical-tensor networks encode these rules using irreducible representations of $SO(3)$, including channels with $L = 0,1,2,\ldots$. Their hidden features therefore retain transformation information that scalar channels do not carry explicitly.

Early implementations of this mathematical framework include Tensor Field Networks (TFN) [86] and Cormorant [87], which demonstrated how rotationally equivariant features can be propagated through tensor-product operations. Subsequently, NequIP [88] established the effectiveness of this approach for interatomic potentials. In architectures such as NequIP, rotating the input produces the corresponding transformation of tensor feature channels through Wigner-D representations. This explicit tensor transformation structure provides a geometric inductive bias for intermediate features and information propagation.

Equivariance can improve data efficiency by encoding the relationship between rotated configurations [88]. Equivariant features also support symmetry-consistent tensorial observables. PaiNN propagates scalar and vector features, with readouts for properties such as polarizability [89]. NewtonNet uses force- and displacement-inspired latent features but obtains physical forces by differentiating its predicted energy [90]. This energy-based construction enforces conservativity for a differentiable model, while numerical energy conservation still depends on the integration protocol.

However, spherical-tensor equivariant networks can incur substantial computational cost. Many architectures combine irreducible representations using Clebsch–Gordan tensor products (Figure 2c), although equivariance can also be implemented through restricted products, Cartesian tensors, local frames, or other symmetry-preserving operators.

Clebsch–Gordan tensor products can contribute substantially to the evaluation cost of equivariant MLIPs. Increasing angular order or channel count generally increases the number of coupling paths and intermediate features. Runtime bottlenecks depend on tensor-product structure, operator implementation, numerical precision, and hardware: some workloads are limited by memory movement, whereas others are limited by arithmetic or kernel-launch overhead. Specialized kernels, fusion, and compilation can improve GPU utilization, as discussed in Section 6. The relevant comparison is therefore measured accuracy and throughput for a specified workload.

Architectural design can reduce these costs through higher-order information within shallow networks or through strictly local interactions. MACE and Allegro illustrate these complementary strategies.

The first pathway uses higher-body-order messages to improve the accuracy–cost trade-off of equivariant architectures. M3GNet and CHGNet demonstrated broad chemical coverage with scalar graph representations, while equivariant message-passing models introduced alternative ways to encode many-body geometry. The MACE architecture, proposed by Batatia et al. [91], addressed this challenge. MACE merges equivariant message passing with the mathematical essence of atomic cluster expansion (ACE). Tensor products of aggregated node features construct higher-body-order local messages, enabling competitive accuracy with two message-passing layers in the reported benchmarks. The spatial receptive field remains determined by the number of layers and interaction cutoffs. This shallow many-body construction provided the architecture for broadly pretrained models such as MACE-MP-0 [22].

Strictly local architectures offer a complementary route to efficient parallel evaluation. Allegro constructs equivariant interactions within local environments without iterative multi-hop node updates [92]. Its locality supports domain decomposition and large atom counts. Chemical coverage remains a separate training choice, while accessible simulation time depends on throughput and the number of integration steps.

A complementary architectural philosophy is illustrated by SUS²-MLIP, developed by Hu et al. [93]. The model incorporates universal-equation-of-state (UEOS) scaling into an element-agnostic radial function, separating chemical identity from the coordinate expansion. The radial-coefficient space therefore loses the explicit $Z^2$ factor associated with element pairs, although element-dependent scaling parameters remain. Nonlinear radial transformations, motivated by associative-memory and KAN ideas, extend the expressive capacity of this compact framework. In the reported multielement benchmarks, SUS²-MLIP achieves accuracy comparable to several neural-network potentials with two to three orders of magnitude fewer parameters; the alloy model has 17,689 parameters, compared with over five million for DPA-2. These benchmarks illustrate the parameter efficiency of physical scaling priors. Parameter count alone does not determine inference speed, memory use, or communication cost; these require measurements at matched accuracy and hardware. On the broader M3GNet dataset, the smaller SUS²-MLIP model has higher errors than M3GNet, revealing an accuracy–compactness trade-off.

These approaches demonstrate that chemical coverage and computational locality are separate design choices. Architecture should be selected together with the required interactions, target observables, and available hardware. Specialization is useful when a deliberate change improves validated performance for those requirements.

## 2.4. Extensions of representations and physical variables

Current developments include Cartesian tensor implementations, equivariant attention, additional electronic-state variables, and physical scaling priors (Figure 2d). These directions address different representational and computational requirements.

Because spherical-tensor implementations can incur costly angular-momentum coupling and irregular memory access, several groups have explored Cartesian equivariant representations. HotPP is an $E(n)$-equivariant message-passing model that propagates Cartesian tensors of arbitrary order; reducible Cartesian tensors preserve equivariance but can contain trace redundancy [94]. TACE develops spectral and spatial tensor atomic cluster expansions in Cartesian space and uses irreducible Cartesian tensor decomposition to organize high-order components; its CG-free spatial formulation provides an alternative implementation route, while high-order tensor evaluation remains a computational cost [95].

TensorNet [96], CACE [97], CAMP [98], and TACE use Cartesian constructions to organize geometric information and symmetry-preserving operations. Tensorial ACE extensions also provide bases for covariant observables. Their accuracy and computational cost depend on tensor rank, body order, basis truncation, implementation, and task.

Alongside these tensor representations, equivariant attention combines symmetry-preserving features with learned information weighting. Its spatial reach depends on which environments are connected: attention restricted by a cutoff remains local.

EquiformerV2 [99] uses efficient $SO(2)$-reduced spherical operations developed in eSCN [100] to support higher-degree local equivariant attention. SO3krates [101] uses Euclidean variables to modulate attention, while eSEN [102] employs edge-aligned $SO(2)$ operations in an equivariant architecture. Studies of approximately learned equivariance provide further comparisons between symmetry constraints, data scale, and task performance [103, 104]. Together, these approaches explore different ways to represent and implement symmetry.

Electronic-state dependence introduces a further representational requirement. Many MLIPs represent one selected adiabatic PES, often for a specified charge and spin treatment. Redox chemistry, magnetic transitions, and nonadiabatic processes can involve additional electronic states. Modeling them may require explicit state variables or multiple coupled PESs, according to the chosen electronic-structure and dynamical description.

Electronic variables can enter as inputs, latent variables, or prediction targets. DeepH predicts DFT Hamiltonian matrix elements for electronic-structure calculations [105]. Spin-dependent symmetry functions distinguish collinear spin environments [106], while fourth-generation HDNNPs use global charge equilibration to represent nonlocal charge transfer [107]. Magnetic ML potentials address spin-dependent energies and spin–lattice coupling [108, 109]. These constructions represent different physical quantities; any nuclear–electronic dynamics must remain consistent with the selected state model.

Architecture, configurational coverage, and reference accuracy limit different aspects of an MLIP. Model refinement should target the dominant limitation for the intended observable and computational budget. Richer representations, targeted sampling, and improved electronic-structure references address model bias, configurational coverage, and label fidelity, respectively.

The Orb-v2 and Orb-v3 family, developed by Rhodes et al., deliberately departs from the dominant equivariant, conservative paradigm. Orb-v3 includes both direct-force and conservative variants based on a non-equivariant architecture and reports near-state-of-the-art accuracy across a range of evaluations, including properties requiring higher-order derivatives of the potential energy surface. Its direct-force variants achieve substantial reductions in latency and memory footprint relative to alternative MLIPs in the reported benchmarks [110]. The authors compare how equivariance, conservativity, and graph sparsity affect accuracy, latency, and memory use. Furthermore, Orb-v3 has demonstrated stable simulations of the carbonic anhydrase II enzyme—a system of over 20,000 atoms that is well outside its training distribution—using orb-v3-direct-inf-omat with Langevin dynamics at 300 K [110]. For NVE dynamics, the authors recommend the continuous conservative-inf variant. A stable thermostatted trajectory does not establish accurate equilibrium statistics, reaction kinetics, or transport. Thermostatting does not automatically remove systematic bias from nonconservative forces [111]. These distinctions motivate selecting and validating the appropriate model variant for the intended observable.

These developments couple representational capacity to computational cost. Targeted exploration addresses gaps in configurational coverage (Section 3), while specialization for demanding applications may also require additional state variables, interactions, or deployment strategies (Section 4).

## 3. Targeted sampling for materials dynamics

As systematically detailed in the preceding sections, the evolution of underlying network architectures—from hand-crafted descriptors to high-order spherical tensors, and toward frontier Cartesian tensors and equivariant Transformers—has pushed the fitting capacity of MLIPs to high levels of physical fidelity. When a model's expressive capacity is no longer the primary bottleneck, the "nature of the data distribution it fits" can become a critical determinant of the force field's performance limits in complex, multi-scale chemical reactions.

### 3.1. Training distributions and PES softening

High-throughput electronic-structure datasets have accelerated MLIP development. Many early materials datasets emphasize stable structures and relaxation trajectories, while molecular datasets often emphasize equilibrium conformers. These resources provide valuable information, but their sampling objectives need not cover the strained, reactive, or interfacial environments visited in a particular simulation.

This emphasis is changing in materials foundation-model development. OMat24 includes extensive off-equilibrium inorganic configurations and reports reduced softening in models trained on its more diverse data [112]. MatterSim uses active learning to sample a wide range of temperatures and pressures [113]. These efforts broaden the coverage of pretrained models; whether additional sampling is needed remains a question for the intended process and observables.

PES softening provides one example of this distribution dependence. Deng et al. reported systematic underestimation of selected vibrational and energetic properties by M3GNet, CHGNet, and MACE-MP-0 [114]. Phonon softening indicates errors in local PES curvature and does not uniquely identify missing high-energy transition-state data as their cause. Surface-energy benchmarks also identified substantial errors for selected elemental surfaces when broadly pretrained MLIPs were applied without task-specific adaptation [115]. Reference quality is a separate constraint: inaccurate electronic-structure labels cannot be corrected by fitting them more closely.

Targeted additions can close coverage gaps in physically constrained models. In $SUS^2$-MLIP, adding 45 FeCo3Sb4Ti4 configurations substantially improved phonon predictions for a previously undersampled Co–Fe interaction [93]. The example shows how focused sampling can complement a physical scaling prior.

For reactive applications, targeted exploration of relevant transition regions and competing pathways can complement broad pretraining and architectural priors. The value of additional data should be tested through barrier estimates, forces along the sampled pathways, and the stability and accuracy of the intended dynamics.

Unbiased MD often revisits accessible basins while rarely crossing high free-energy barriers on available timescales. Heating can broaden exploration, but it need not reproduce the pathways or environments relevant at operating conditions. Descriptor-based metadynamics can access configurations missed by conventional MD (Figure 3b) [116]. Comparisons with saddle-search methods likewise show that targeted exploration can provide complementary training structures [117].

Targeted sampling should complement equilibrium data rather than replace it. A model used for reactions or deformation must describe both the transition regions and the basins that determine populations and relaxation. The following methods provide different ways to obtain that coverage.

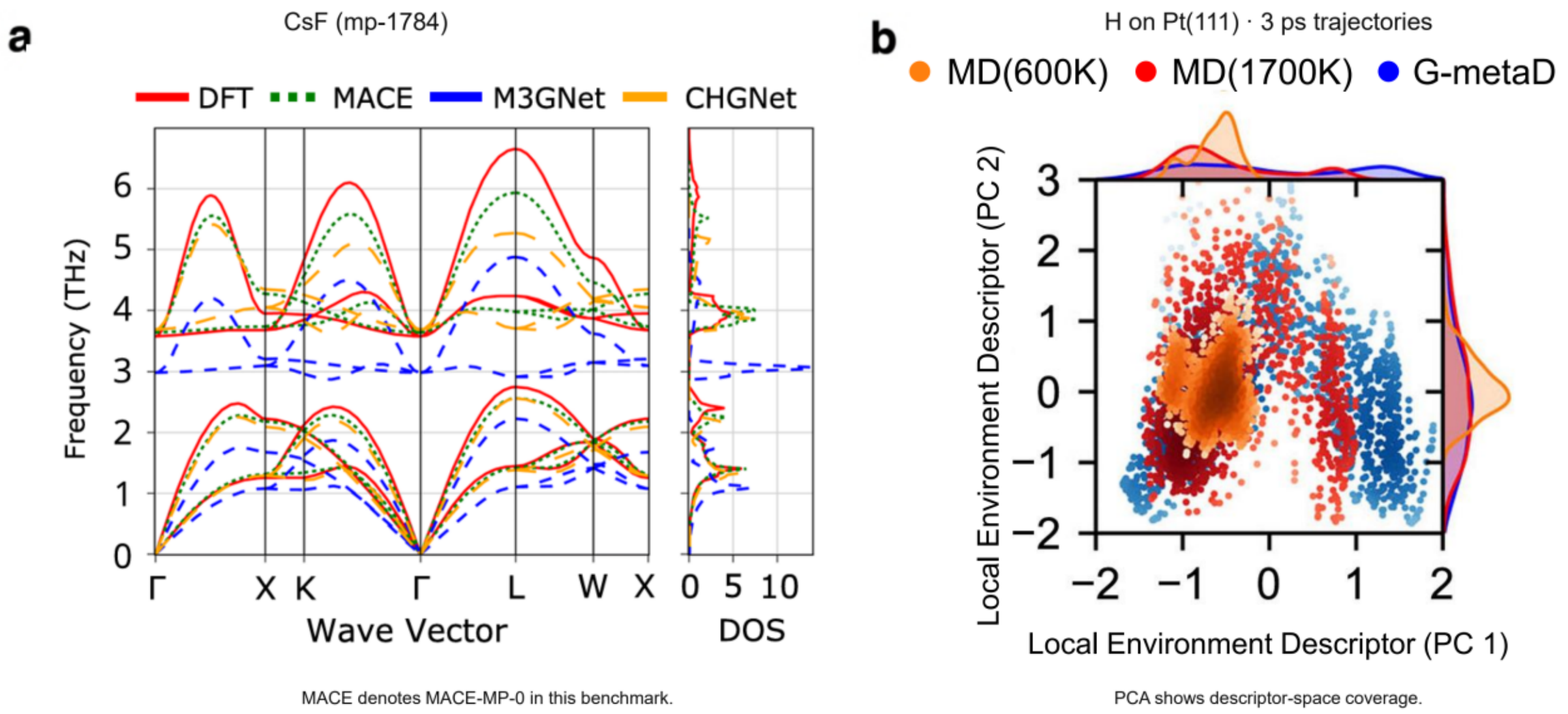

Figure 3. Potential energy surface softening in tested foundation models and complementary targeted exploration. (a) Phonon dispersion relations and density of states (DOS) for CsF (mp-1784). The solid red line is the density-functional-theory (DFT) reference; dashed and dotted lines are predictions from the tested universal machine-learning interatomic potentials, including MACE-MP-0 (labeled MACE), M3GNet, and CHGNet. In this benchmark, several models underestimate parts of the vibrational spectrum, indicating errors in local PES curvature consistent with softening. Adapted from Ref. [114]. (b) Principal component analysis (PCA) of the local atomic environment descriptors explored during different atomistic sampling strategies. For H on Pt(111), the illustrated 3 ps trajectories compare MD at 600 K (orange dots) and 1700 K (red dots) with metadynamics using a local environment descriptor as a collective variable (G-metaD, blue dots). G-metaD explores a broader region of descriptor space than the conventional MD trajectories in this example. The PCA projection compares configurational coverage in descriptor space. Targeted exploration can provide additional reference configurations for subsequent dynamical validation. Adapted from Fig. 2e of Ref. [116].

### 3.2. Targeted sampling and active learning

Active learning prioritizes configurations for reference labeling, but it does not uniquely determine how those configurations are generated. An exploration trajectory may remain trapped in one basin even when its uncertainty is small. Conversely, large model disagreement can arise in physically irrelevant regions. Combining uncertainty indicators with physically motivated exploration helps focus labeling on the intended problem. Reaction endpoints, local curvature, collective variables, and chemical rules provide complementary information for exploration. Their usefulness depends on what is known about the process and which states remain undersampled.

#### *3.2.1. Paths between known endpoints*

When the initial and final structures of a reaction are known, the Nudged Elastic Band (NEB) method is a widely used approach for locating a minimum-energy path. The NEB and climbing-image NEB (CI-NEB) algorithms [118, 119] construct a chain of images connected by springs between reactants and products. The nudging projection removes the perpendicular spring component and the parallel component of the true force; the climbing image additionally ascends along the path to approach a first-order saddle point. Convergence and pathway quality still depend on the initial path, number of images, and underlying potential-energy surface.

Pathway datasets constrain configurations between stable reactants and products. OC20NEB provides catalytic reaction paths [120], while Transition1x samples organic reaction pathways [121]. NeuralNEB uses learned potentials to accelerate path optimization [122]. Reinforcement-learning approaches also explore transition states and minimum-energy paths [123]. These methods accelerate pathway searches, with candidate paths assessed against the chosen electronic-structure reference.

### *3.2.2. Curvature-guided and endpoint-free exploration*

When products are unknown, endpoint-based path optimization cannot define the search by itself. Local curvature and stochastic exploration can instead identify neighboring basins or candidate saddle points. The resulting configurations complement trajectories confined to a known reaction channel.

Normal-mode sampling generates displaced structures from harmonic vibrational modes near a stationary point. It contributed to ANI development [63], and Hessian-informed data can improve curvature-related predictions [124]. Related strategies appear in ANI-1x [125], ANI-2x [65], QM7-X [126], and organometallic reaction workflows [127]. This efficiently broadens local sampling, although strongly anharmonic regions and new reaction channels may require complementary exploration.

The dimer method searches for a nearby first-order saddle using forces to approximate the lowest-curvature direction [128]. Waters and Rondinelli compared such structural-evolution strategies for constructing MLIP training sets in Zr–O systems [117]. Their results support combining methods that sample different regions of configuration space. Activation-relaxation techniques, including r-ARTn and d-ARTn, provide related endpoint-free searches [129].

Stochastic surface walking (SSW) combines biased displacement and local relaxation to explore basins and pathways [130]. Coupling SSW with learned potentials has supported global searches of catalytic reaction networks, including ethene epoxidation [131]. LASP integrates these components in an automated workflow [132]. Sampling remains dependent on the explored composition, boundary conditions, and quality of the underlying potential.

### *3.2.3. Biased dynamics and collective variables*

Biased dynamics uses collective variables (CVs), such as bond distances or coordination numbers, to promote otherwise infrequent events. The chosen variables determine which pathways are accelerated. Because the bias changes the sampled dynamics, extracting equilibrium probabilities or kinetic rates requires the corresponding statistical treatment.

Constraints and biasing forces serve different roles in molecular dynamics. SHAKE enforces holonomic constraints during integration [133]. Restrained or steered simulations instead apply forces that move selected coordinates. Zhang et al. generated ANI-1xBB using bond-breaking-oriented sampling [134], while Hong et al. used a sum of bond distances as a collective variable in steered MD for HF dissociative adsorption on silicon nitride [135].

More advanced methods include umbrella sampling and metadynamics. Umbrella sampling [136] applies biased windows along selected collective variables and can reconstruct a free-energy profile after appropriate reweighting; it has been combined with MLIPs for atomic-layer-deposition and catalytic applications [137, 138]. Metadynamics [139] deposits a history-dependent bias in collective-variable space to accelerate escape from visited basins. Free energies can be recovered under method-specific convergence and reweighting conditions, whereas unbiased kinetic rates generally require additional treatment. Descriptor-space metadynamics has been used to seek undersampled atomic environments [116]; CREST uses metadynamics-like biases for conformer exploration [140]. ArcaNN combines concurrent learning with OPES for reactive pathways [141], and restricted multicanonical sampling has been used to broaden energy-space exploration [142].

### *3.2.4. Chemical rules and reaction-network enumeration*

Rule-based enumeration offers another route when direct trajectories would explore too few reaction channels. Proposed bond rearrangements generate candidate reactions that can subsequently be screened and refined using electronic-structure calculations. The imposed rules also define which chemistry may be missed.

YARP automates reaction proposals and transition-state searches using chemical transformation rules [143]. The associated RGD1 dataset provides an extensive collection of elementary organic reactions [144]. These resources can expand reactive training coverage, but enumeration is bounded by the allowed transformations and search settings. Applicability to metals or extended solids requires different structural and chemical assumptions.

### *3.2.5. Concurrent learning and defect-focused sampling*

Concurrent learning links exploration, configuration selection, reference labeling, and model updates. Physical conditions guide exploration, while model-dependent indicators help allocate labeling effort. The challenge is to cover relevant events without repeatedly sampling similar environments.

DP-GEN (Deep Potential Generator) established a widely used concurrent-learning workflow for Deep Potential models [145]. During exploration under selected temperatures, pressures, and other boundary conditions, an ensemble of independently trained models provides a model-deviation score that is used to select configurations for additional quantum-mechanical labeling. Ensemble disagreement serves as an extrapolation indicator whose sensitivity depends on model diversity and configurational coverage. Agreement can persist outside the training domain. Zhu et al. estimated uncertainty with a single neural network, reducing the overhead of ensemble-based uncertainty evaluation [146]. Kulichenko et al. used an uncertainty-derived bias to drive exploration toward underrepresented configurations [147]. Dataset-selection methods seek compact informative subsets [148]. Related workflows couple reaction rules with active learning (e.g., RAL) [149], use nanoreactors for exploratory sampling [150], or combine enhanced sampling with active learning at catalytic interfaces (e.g., DEAL) [151].

Extended metallic defects introduce an additional sampling problem: their characteristic dimensions can exceed a tractable DFT cell. Shuang et al. sampled defect environments in large empirical-potential simulations and reconstructed selected clusters as periodic reference cells [152]. MLIPs trained with this approach were tested in tungsten nanoindentation, tensile deformation, and fracture. The example shows how physical sampling and reference-cell construction can make dislocation networks and grain-boundary environments accessible to first-principles labeling.

For hematite–water interfaces, Schienbein et al. used active learning to obtain a potential for nanosecond-scale simulations [153]. Combustion studies likewise require coverage of multiple intermediates and reaction channels [154, 155]. Yoo et al. used ReaxFF trajectories as exploration seeds and relabeled selected configurations with quantum calculations [156]. These workflows connect exploratory reactive trajectories to quantum-mechanical reference labels.

Architecture and sampling constrain different aspects of a learned potential. Path searches, normal-mode displacements, biased trajectories, reaction rules, and defect reconstruction provide complementary configurations. Their combination should be chosen for the mechanisms and observables of interest, which motivates the application-dependent strategies considered next.

## 4. Application requirements and physics-driven specialization

Through the preceding discussion of architecture and sampling, a central distinction emerges between broad chemical coverage and validated fidelity under specified conditions. Universal models provide useful zero-shot baselines, but their reliability does not automatically extend to new ensembles, external fields, reactive pathways, or length and time scales. This section examines four scenarios in which specialized data, physical variables, or simulation algorithms are often required.

Here, specialization denotes a deliberate adjustment of model inputs, representation, reference data, training, or deployment for a specified system, physical condition, or observable. Its benefit should be assessed through the intended observables and against a suitable baseline, including the unadapted pretrained model where task-compatible. General choices made when building a broadly applicable foundation model do not, on their own, constitute downstream specialization. MACE-MP-0, for example, reports direct applications across diverse systems as well as routes to further fine-tuning [22].

### 4.1. Foundation models and routes to adaptation

With the expansion of large-scale first-principles databases (e.g., Materials Project, OQMD), machine learning potentials are undergoing a paradigm shift toward "Foundation Models." Universal models represented by M3GNet, CHGNet, and MACE-MP-0, trained largely on structures from ground-state relaxation datasets, have demonstrated useful zero-shot generalization potential, significantly enhancing the screening throughput for near-equilibrium properties (e.g., lattice constants, elastic tensors, formation energies, and phonon spectra). Multi-domain pretraining has expanded this approach. JMP uses a GemNet-OC directional message-passing backbone for joint pretraining on molecular and catalyst datasets, followed by downstream fine-tuning, including materials tasks [157]. Its pretraining force head is equivariant, whereas its backbone uses scalar node and edge features rather than non-scalar spherical-tensor features. In the Deep Potential Attention (DPA) series, DPA-1/2 use attention-based descriptors [158, 159], whereas DPA-3 operates on invariant features within a line graph series [160].

Foundation models provide useful zero-shot baselines across broad chemical spaces, especially for properties close to the distributions represented in their training data. Reliability in reactive, high-energy, charged, or externally driven states depends on how these environments are represented in the training data and model inputs.

DPA-3 illustrates how representation and training can be designed for broad transferability [160]. Its line graph series (LiGS) framework propagates invariant node and edge features and increases model capacity through layer stacking. Dataset-aware multi-task training supports heterogeneous reference data within a shared fitting framework. The DPA-3.1-3M model trained on OpenLAM-v1 demonstrated zero-shot force-field prediction across 12 downstream tasks. These benchmarks assess energy, force, or related prediction errors on the evaluated datasets; validation of a particular reaction trajectory or kinetic observable remains a separate assessment.

To improve fidelity relative to higher-level electronic-structure references, universal foundation models and their derivative frameworks have extensively incorporated multi-fidelity and delta learning ($\Delta$-learning). In $\Delta$-learning, models explicitly fit differences between a lower-cost baseline and a higher-level reference, reducing the number of expensive reference calculations needed [161, 162]. Multi-fidelity models can instead share representations while conditioning predictions on reference level; SevenNet-MF uses fidelity-dependent parameters to learn from abundant lower-fidelity and sparse higher-fidelity data [163]. MACE-MH uses multi-head replay post-training across electronic-structure theories and chemical domains, preserving information from earlier tasks while incorporating new ones [164]. These tasks need not form a single low-to-high accuracy hierarchy. To accommodate heterogeneous training tasks, UMA uses a Mixture of Linear Experts (MoLE), with routing conditioned on composition and global context, including charge, spin, and the target DFT task [165]. The routing distinguishes predictions associated with different reference tasks. Multitask and multi-fidelity strategies can improve data efficiency by combining information across reference levels [166]. DPA-3 conditions a shared fitting framework on one-hot dataset-identity encodings to accommodate heterogeneous electronic-structure settings [160]. Knowledge distillation uses teacher-model predictions as training targets for a student potential, as demonstrated for organic molecular crystals [167].

Multi-fidelity learning can improve accuracy relative to a chosen high-level reference while reducing the number of expensive labels [168]. The attainable fidelity and transferability nevertheless depend on the high-level configurations used for correction. Methods such as canonical CCSD(T) have steep formal scaling (commonly quoted as $\mathcal{O}(N^7)$) and are usually limited to comparatively small systems, so reactive and high-energy configurations may remain sparsely labeled. Including reactive configurations in the correction set extends the scope of the higher-level description.

The MACE-OFF family targets organic molecular modeling: trained on the SPICE dataset and extended in the MACE-OFF24-SC model with softcore interactions and message scaling to enable alchemical free energy calculations [169], MACE-OFF achieves subchemical accuracy relative to experiment for solvation free energies and demonstrates robust behavior in challenging organic crystal structure prediction tasks. These results support the use of chemical-domain models for the tested molecular and free-energy applications [169, 170]. Chemical-domain pretraining and the downstream softcore modifications are distinct design choices. Meta FAIR's Open Molecules 2025 (OMol25) dataset broadens molecular reference coverage. OMol25 comprises over 100 million DFT calculations at the ωB97M-V/def2-TZVPD level, explicitly including “reactive structures” and “variable charge/spin” alongside conformers and solvated systems [171]. A recent preprint reports 96% transition-state search success rates for OMol25-trained models (MACE-OMol25, UMA-M) on the Poly25 closed-shell polymerization benchmark using the native freezing-string workflow followed by DFT transition-state optimization [172]. The reported search result demonstrates the value of reactive reference data within that workflow.

For reactive MD, validation should include the bond rearrangements, short-range encounters, and competing intermediates visited under the target conditions.

Short atomistic trajectories resolve local rearrangements and early-stage structural evolution. Simulations of lithium–electrolyte interfaces, for example, have proposed nanoscopic precursors to dendrite initiation [173]. Connecting these precursors to macroscopic growth requires longer-time kinetic modeling.

Reference accuracy and configurational coverage address different limitations. Higher-level labels improve the target PES only where the reference method is appropriate and the relevant configurations are represented. Neither additional labels nor longer trajectories alone establish reliable phase stability or reaction kinetics.

Two common routes to specialization are fine-tuning a pretrained model and training a system-specific potential. The choice depends on data availability, reference compatibility, physical variables, and the accuracy–cost target.

Fine-tuning can reuse a broad representation for complex mixtures or multicomponent alloys. Its value depends on compatibility between the pretrained reference and the new labels, as well as coverage of the target configurations. For a narrower domain with stringent throughput requirements, a compact model trained from scratch can be competitive. It forfeits pretrained information but allows direct control of the representation and dataset. Matched comparisons should evaluate accuracy, dynamical observables, retention of required behavior, and evaluation cost.

Architectural priors provide a complementary choice within either fine-tuning or training from scratch. The physical scaling example in Section 2.3 illustrates how representation, data coverage, and parameter efficiency can be considered together.

Retaining useful pretrained behavior is an additional consideration during fine-tuning. Lifelong-learning and replay strategies can reduce forgetting [174], while transferable models can support related applications across chemical space [175]. Forgetting-aware fine-tuning methods, including reEWC, explicitly balance adaptation and retention [176]. Their success should be assessed on both the new target and the regions that the adapted potential must continue to represent.

The following scenarios organize these choices by physical requirement. They illustrate where specialization has been useful and which tests are needed to assess its value. Four settings are considered: potential-controlled interfaces, particle-exchanging systems, multiphase evolution, and large-scale deformation. These settings overlap and include equilibrium and driven processes. Constant-potential or reservoir sampling need not be nonequilibrium, and quasistatic fracture calculations do not provide an MD time axis. Each combines distinct demands on the potential and the simulation method. Metallic applications are included throughout because phase competition, evolving defects, and chemical order provide stringent tests of dynamical reliability.

At electrochemical interfaces, the model must represent the response to potential, charge, and external fields. For open systems, the potential must support changing composition and the configurations generated by deposition, insertion, removal, or irradiation. For multiphase evolution, competing free energies, interface structures, and activated processes must be represented at the relevant conditions. For deformation and fracture, defect energetics must remain reliable at the length and time scales needed to resolve collective mechanisms. Figure 4 gives representative examples of these complementary requirements.

### 4.2. Electrochemical interfaces and constant-potential simulations

Many foundation MLIPs are energy models conditioned on atomic species and geometry at a fixed electronic reference state. Unless charge, electrode potential, or an external field is supplied explicitly, such models cannot distinguish configurations that share the same atomic geometry but correspond to different electrochemical boundary conditions.

Electrochemical experiments are often controlled at an electrode potential. In an ideal constant-potential description, the electrode can exchange electrons with a reservoir so that the electronic chemical potential is fixed.

The learned thermodynamic potential must also match the electronic ensemble. In an adiabatic constant-potential description with equilibrated electrons, the relevant quantity is $\Omega = E - \mu_e N_e$, where $\mu_e$ is the electronic chemical potential and $N_e$ is the electron number relative to a consistent reference. Atomic forces follow the negative derivative of Ω with respect to nuclear coordinates at fixed $\mu_e$, rather than that of the canonical energy E alone [177]. At finite electronic temperature, the corresponding electronic free energy replaces E in this transformation.

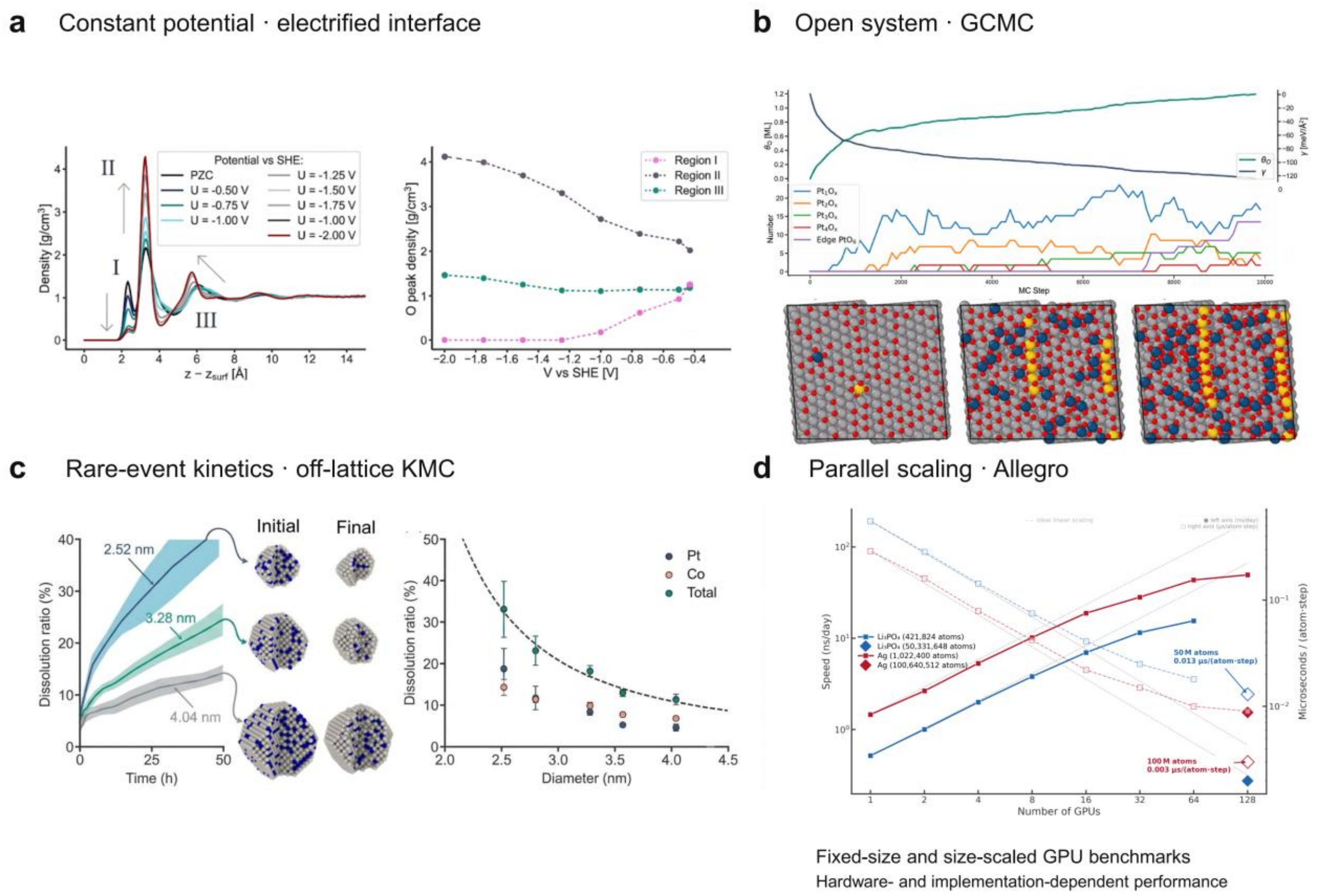

Figure 4. Representative application requirements and MLIP-enabled workflows. (a) Field-coupled ensemble (Constant Potential). Density profiles (left) and potential-dependent oxygen peak densities (right) of water molecules at the electrified interface, showing potential-dependent changes in interfacial water structure across separately sampled constant-potential conditions ($U$). Adapted with permission from Ref. [178]. (b) Particle exchange and composition changes (GCMC). Evolution of oxygen coverage ($\theta_O$), surface formation energy ($\gamma$), and statistics of surface oxides during the oxidation of a stepped Pt surface, coupled with representative 3D atomic snapshots at different oxidation stages. Adapted with permission from Ref. [179]. (c) Macro-time dynamic evolution (KMC). Simulated dissolution ratios of $Pt_3Co$ nanoparticles over macroscopic time scales (up to 50 hours) and the size-dependent degradation trends captured by off-lattice kinetic Monte Carlo driven by a specialized MLIP. Adapted with permission from Ref. [180]. (d) Macro-space parallel scalability. Fixed-size and size-scaled GPU benchmarks for the local equivariant Allegro architecture, including a reported simulation of 100 million atoms on 128 GPUs. Performance depends on hardware, system density, precision, and implementation. Adapted with permission from Ref. [92].

Field- and potential-conditioned models address this limitation by including the variables that distinguish the relevant electronic states. Their construction must also specify how those variables couple to atomic motion.

For field-dependent models, external forces can be added consistently when their coupling is known, but a learned field-dependent energy must transform correctly with both the atomic coordinates and the field. Because an electric field is a polar vector, the relevant transformation rules are those of $E(3)$, including inversion parity, rather than rotations alone. Zhang and Jiang introduced pseudoatomic field vectors to construct field-aware equivariant descriptors [181]. Falletta et al. instead derived polarization and Born effective charges as derivatives of a generalized energy with respect to an external field [182]. These constructions relate the modeled energy directly to field-dependent response properties.

For constant-potential simulations, Wang et al. developed CP-MLFF within the MACE framework, using electron number as an input and predicting the Fermi level [183]. Bianchi et al. developed TRECI, combining transferred shared descriptors with potential-specific readouts to reproduce potential-dependent structural and dynamical properties of the Cu/water interface (Figure 4a) [178]. Related field-aware models can describe drift and dielectric response in ionic systems when the electrostatic and field couplings are represented consistently [184].

In battery systems, interphase formation and component diffusion can involve redox chemistry and configurations far from the training distributions of many current universal models. Targeted sampling and system-specific training can improve their representation.

Ren et al. trained a force field for redox evolution at a lithium-metal/solid-electrolyte interface and examined atomistic interphase formation [185]. Fu et al. developed a reactive MLIP for Li–Si anodes to study heterogeneous interdiffusion [186]. Both applications connect representative reactive configurations to a defined interfacial process.

Electrocatalytic interfaces combine field response, solvation, structural fluctuations, and activated chemical events. Uncertainty-guided labeling helps select configurations, while enhanced sampling can promote events that unbiased trajectories rarely visit. Their combination is useful when both configurational coverage and labeling cost limit model development.

For oxygen reduction at an Au–water interface, Yang et al. combined neural-network potentials with metadynamics to sample reactive configurations [187]. Raman et al. used iterative learning to investigate acid–base chemistry at an $IrO_2$ interface [188]. Nakanishi et al. examined proton transfer at a $Zr_7O_8N_4$/water interface with a model adapted to its surface environments [189]. These examples link model construction to specific interfacial processes.

Electrochemical simulations require a consistent description of potential, charge, field, and sampling conditions. Task-specific training and broadly pretrained conditional models can both address these requirements. Their suitability should be judged by potential-dependent structures, energetics, and response functions.

### 4.3. Open systems and changing composition

Open-system simulations introduce a different requirement: exchange of atoms with an environment. The potential must evaluate the resulting compositions and local structures, while the sampling algorithm determines how exchange occurs.

Reliable open-system energies require consistent references across compositions and coverage of the local environments generated by particle exchange. Open-system processes such as CVD, ALD, plasma etching, and particle implantation involve particle exchange or imposed reactive fluxes and can generate short-range collisions, dangling bonds, interstitial defects, and distorted coordinations. Models not trained or regularized for these configurations may produce unphysical energies or forces and destabilize a trajectory.

Useful workflows combine variable-composition potentials with reservoir sampling, explicit deposition, or reactive dynamics. Training and short-range regularization must cover the environments generated by the chosen protocol.

Grand-canonical sampling describes equilibrium with a reservoir at a specified chemical potential ($\mu$). It is useful for composition-dependent surface structures and oxidation states. Driven growth or reactive fluxes may instead require explicitly non-equilibrium protocols.

Conventional fixed-particle MD can sample NVE, NVT, or NPT conditions but does not by itself exchange atoms with a reservoir. Xu et al. coupled an Embedded Atom Neural Network (EANN) with grand-canonical Monte Carlo (GCMC) to sample variable composition at a specified chemical potential [179].

GCMC evaluates energy differences between states with $N$ and $N \pm 1$ particles, with acceptance probabilities chosen for the intended reservoir conditions. The potential must give comparable energies across those states. Using EANN energies, Xu et al. examined oxidation-induced coarsening and step reconstruction of Pt surfaces (Figure 4b) [179]. This coupling samples equilibrium structures across particle numbers, expanding the accessible composition space.

Deposition and implantation generate energetic collisions and unusual local environments. Caro et al. simulated amorphous-carbon deposition using an existing GAP and used Gaussian-process uncertainty estimates to assess the representativeness of local environments encountered during atomic impacts [37]. Cusentino et al. used a linear SNAP potential to investigate helium implantation and suppressed bubble nucleation in $WBe_2$ [190]. These examples link model development to process-specific configurations and defect outcomes.

Chemical etching and atomic layer deposition (ALD) involve bond rearrangements that can be rare on direct MD timescales. For HF reactions with amorphous $Si_3N_4$, Hong et al. used bond-related CVs and enhanced sampling to explore dissociative adsorption and network disruption [135]. Nakata et al. combined learned potentials with umbrella sampling for precursor reactions on silicon surfaces [137]. These approaches target specific reaction coordinates. Free-energy reconstruction and kinetic interpretation require appropriate treatment of the applied bias.

Open-system reliability depends on both the potential and the exchange protocol. Reservoir sampling, explicit impacts, and biased reactive trajectories generate different configuration distributions. Validation should therefore include the short-range interactions, composition changes, and reaction channels relevant to the selected workflow.

### 4.4. Phase transformations and multiphase evolution

Phase transformations and multiphase evolution place simultaneous demands on thermodynamic accuracy and sampling. Competing phases can differ by small free energies, while nucleation, diffusion, and interface migration may occur on very different timescales.

Multiphase sintering, metallic phase transitions, and catalyst aging can involve coarsening, interdiffusion, phase-boundary migration, and dissolution over timescales far beyond direct atomistic MD. The practical time accessible to an MLIP-driven trajectory depends on system size, hardware, and model cost; even when nanoseconds or longer are feasible, hour- or month-scale rare events generally require accelerated dynamics or coarse-grained kinetic models. Phase boundaries can also be sensitive to small free-energy differences, so a model trained to a particular DFT approximation inherits its systematic reference error unless corrected to a higher-level target.

Accelerated sampling and higher-level corrections address different parts of this problem. Kinetic models can extend accessible times, while multifidelity or $\Delta$-learning approaches can refine the reference energetics.

To extend kinetic sampling, Jung et al. coupled a trained MLIP to off-lattice kinetic Monte Carlo to study $Pt_3Co$ nanoparticle degradation [180]. The potential supplied energetics for activated events, enabling predictions over times beyond direct MD (Figure 4c). Such simulations can connect vacancy processes to nanoparticle dissolution, but their temporal interpretation depends on the event catalog, barrier estimates, and rate assumptions.

Sumaria et al. combined a learned potential with basin-hopping Monte Carlo to search catalyst structures under reactive gas conditions [191]. Basin hopping explores low-energy configurations by alternating perturbation and minimization. It can identify structural candidates missed by a short trajectory, but its search sequence does not specify a time-resolved reconstruction mechanism.

Metallic transformations couple lattice rearrangements to composition and chemical order. In multicomponent alloys, these dependencies demand validation across both structural and chemical environments. Kostiuchenko et al. combined an on-lattice low-rank potential (LRP) with Monte Carlo simulations to examine the influence of local lattice relaxation on phase transitions in bcc NbMoTaW [192]. Local relaxation effects were incorporated implicitly by fitting energies of DFT-relaxed configurations. The resulting phase behavior depends on composition, temperature, and chemical arrangement.

Accurate sampling alone is insufficient if the reference misrepresents phase competition. Solid-solid transition temperatures may depend on small free-energy differences and can be sensitive to the electronic-structure reference, sampling, and nuclear quantum effects. Models trained to GGA-level data can therefore inherit material-dependent transition-temperature errors.

Liu et al. used a $\Delta$-learning strategy for the high-temperature phase transition of zirconia [193]. A lower-level DFT model was used for broad sampling, and selected configurations were labeled with the random-phase approximation (RPA); the learned energy correction improved agreement with the experimental transition behavior at substantially lower cost than labeling the full dataset at the RPA level.

Phase-evolution studies therefore need to distinguish pathway discovery, equilibrium sampling, and kinetic prediction. Direct MD, accelerated methods, and higher-level energy corrections serve different purposes. Their combination is useful when each component is validated for the transformation being modeled.

### 4.5. Deformation, defect nucleation, and fracture across scales

Plastic deformation and failure are collective processes. Their simulation requires both sufficient system size and reliable energetics for the defects that carry strain or initiate damage.

Recent metallic applications show how improvements in sampling and model efficiency can support mechanistic analysis. Figure 5 summarizes representative mechanisms and the distinct simulation protocols used to resolve them.

Dislocation motion, grain-boundary migration, and fracture may involve large defect networks and repeated local events. Computational cost limits the length and duration of a trajectory, while training coverage limits its physical reliability. Increasing the simulated volume is therefore most useful when competing deformation pathways remain accurately resolved.

Li et al. combined entropy–symmetry-guided polymorph generation with knowledge distillation to construct efficient potentials for metal plasticity [194]. MD of iridium nanopillars linked compression to stacking-fault formation and twin propagation, whereas aluminum showed transient twinning, detwinning, and subsequent dislocation slip. The contrasting pathways connect the sampled polymorphic landscape to competing deformation mechanisms. Distillation reduced evaluation cost while retaining the teacher model's relevant information.

Defect nucleation can require validation of free-energy barriers rather than force errors alone. Allera et al. combined dislocation-trained MLIPs for BCC Fe and W with constrained MD and projected-average-force integration to obtain activation free energies for kink-pair nucleation [195]. The resulting activation entropy varied weakly with stress and temperature over the investigated regimes, unlike the strong variations obtained with some empirical potentials. Direct MD provided an additional check through dislocation velocities, while a thermally activated model reproduced experimental yield-stress trends. This connects learned dislocation energetics to activation thermodynamics and the resulting plastic response.

Hydrogen can instead couple defect evolution to embrittlement. Bao et al. developed a neuroevolution potential for W–H and used MD to study hydrogen accumulation near voids and the resulting tensile response [196]. Hydrogen-rich planar structures were associated with suppressed dislocation activity and enhanced cracking on {100} planes under the tested conditions. The mechanistic interpretation links local hydrogen organization to a change in the competition between plastic accommodation and cleavage.

A complementary, quasi-static view comes from Wang et al., who used MLIPs to compare crack extension in Nb, NbMoTaW, and Nb45Ta25Ti15Hf15 [197]. Atomic-scale resistance curves revealed changes in fracture mode as cracks advanced, while local chemical fluctuations affected cleavage, twin nucleation, and dislocation emission. The more ductile Nb–Ta–Ti–Hf alloy exhibited greater crack-tip blunting and resistance to propagation. Incremental loading and energy minimization resolve the competition between these crack-tip mechanisms in the quasi-static limit.

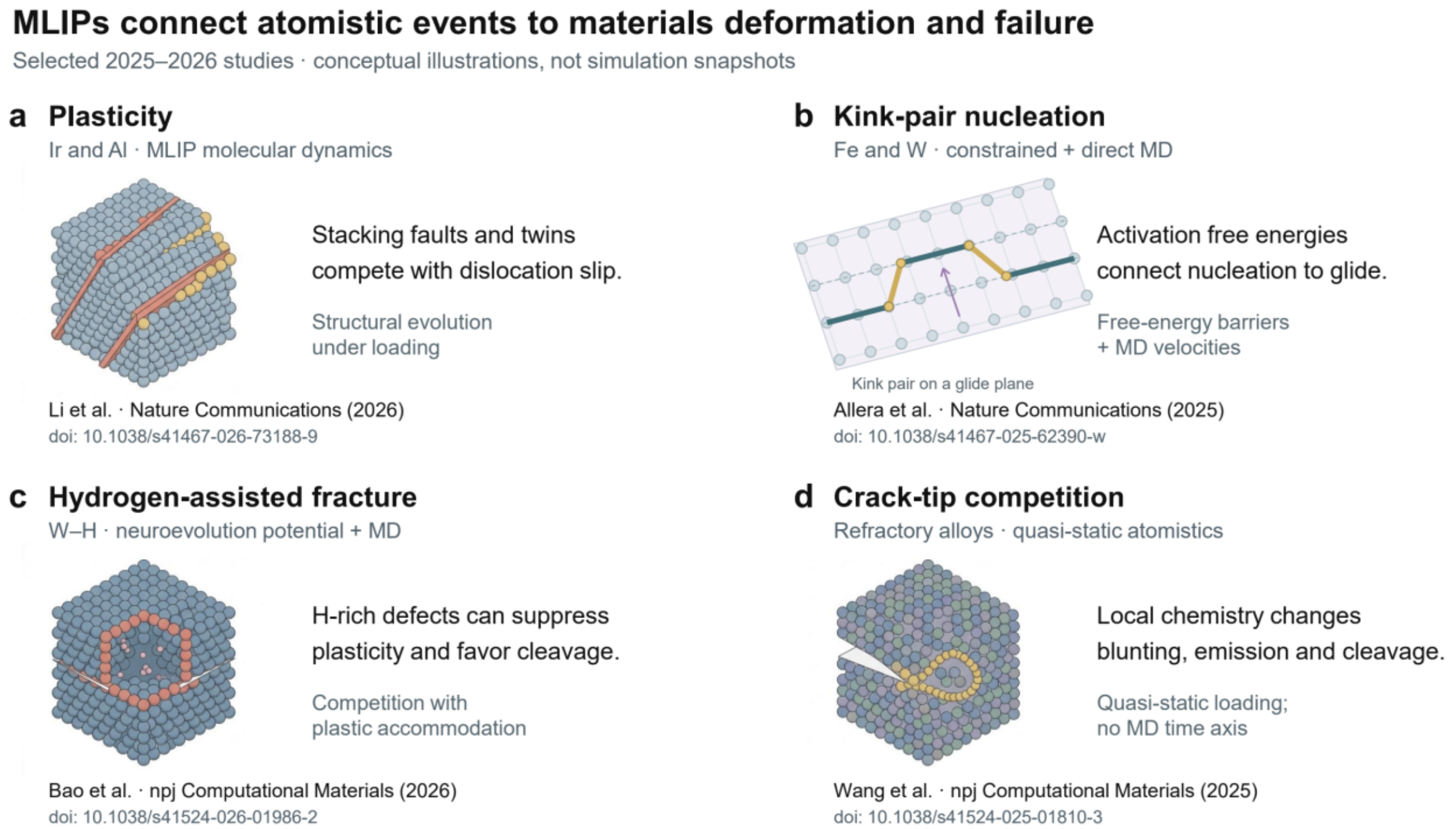

Figure 5. Mechanistic insights into materials deformation and failure from specialized MLIPs. (a) Competing plasticity pathways. MLIP-driven MD of Ir and Al connects stacking-fault formation, twinning, and dislocation slip to the sampled polymorphic landscape [194]. (b) Kink-pair nucleation. For BCC Fe and W, constrained MD and projected-average-force integration resolve activation free energies, while direct MD provides a comparison through dislocation velocities [195]. (c) Hydrogen-assisted fracture. W–H simulations link hydrogen-rich defect structures to suppressed dislocation activity and enhanced cleavage under the tested loading conditions [196]. (d) Crack-tip competition. Quasi-static simulations of refractory alloys connect local chemical environments to crack-tip blunting, dislocation emission, and cleavage under incremental loading and energy minimization [197]. The panels are qualitative mechanism schematics; atom colors and positions are illustrative.

Resolving such collective defect processes at scale also creates computational requirements. Large-scale MD commonly uses spatial domain decomposition. An $L$-layer message-passing model with per-layer cutoff $R_c$ has an effective receptive field up to approximately $LR_c$. Distributed evaluation must communicate the required information across subdomain boundaries. The associated cost depends on partitioning, density, geometry, and whether messages are exchanged layer by layer.

Software–hardware co-design has substantially increased the scale of MLIP-driven MD. Optimized DeePMD-kit operators, memory handling, and distributed integration enabled simulations of 100 million atoms in 2020 [19, 72]. This is a computational scaling milestone; the physical length scale depends on the simulation-cell dimensions and geometry, and larger-scale predictions require additional modeling and validation. NEP/GPUMD provides a complementary deployment route based on efficient local descriptors and GPU execution [68, 69], illustrated here by the W–H defect and fracture application [196]. Evaluation cost, communication, and the target physical observables should be compared together.

Distributed implementations reduce communication and redundant work in different ways. DistMLIP partitions the atomic graph to distribute message-passing operations [198]. For REANN, Xia and Jiang developed forward and backward synchronization across subdomains, enabling efficient parallel force evaluation [199]. These approaches improve throughput within their tested architectures and hardware. Their performance should be compared at matched system sizes, accuracy, and communication conditions.

The chemtrain-deploy framework combines graph pruning, fixed-shape buffers, and deployment optimizations for models including Allegro and MACE in LAMMPS [200]. Based on graph atomic cluster expansion, GRACE-FS supports efficient CPU execution [201, 202]. These developments make hardware and evaluation strategy part of the accuracy–cost comparison. Explicit nonlocal physics, when needed, adds further computational and communication requirements.

Spatial scalability must be accompanied by reliable time integration. Long-time integration also requires separating numerical drift from model error. Energy-derived MLIPs are conservative up to numerical differentiation and floating-point effects, whereas arbitrary direct-force models need not be. Even a conservative potential can enter poorly sampled regions and predict incorrect forces or phase stability. Time-step choice, integrator, thermostat, precision, and model extrapolation should therefore be tested independently.

Across these settings, specialization connects model design to the observables needed to resolve materials mechanisms. Table 1 organizes this assessment: first establish the required physical variables and energy references, then examine configurational coverage, and finally test the observables supporting the intended scientific claim. Model adaptation and changes to the sampling algorithm address different requirements and should be assessed separately.

Table 1. Application requirements and validation priorities for selecting or adapting an MLIP. The framework links the cases in Section 4 to tests of the intended observables. The contexts overlap, and appropriate tests and tolerances depend on the application.

| **Application context** | **Modeling conditions** | **Priority validation observables** |
|---|---|---|
| Constant-potential interfaces [177, 178] | Potential reference; electron exchange or potential conditioning; thermodynamic potential and consistent forces. | Potential-dependent interfacial structure and charging response; target reaction energetics or free-energy changes. |
| Particle exchange and open systems [179, 190] | Consistent elemental energy references and chemical potentials; exchange protocol; coverage of collisions and undercoordinated environments. | Composition-change, adsorption/desorption and defect energies; short-range repulsion; coverage and structure under the chosen protocol. |
| Phase transformations and multiphase evolution [180, 192] | Coverage of competing phases and interfaces; free-energy accuracy; assumptions connecting sampling to kinetics. | Relative free energies and coexistence; transformation behavior; diffusion, interface migration and relevant activation free energies. |
| Deformation and fracture [194, 195, 196, 197] | Coverage of defects and surfaces; loading conditions; finite-size effects; deployment at the required scale. | Generalized stacking-fault energies; dislocation cores and migration; work of separation; competing deformation mechanisms; throughput at matched accuracy. |

## 5. Physical consistency and validation

Throughout the preceding four sections, we have seen how architectural evolution and rigorous sampling have propelled machine learning interatomic potentials (MLIPs) to the forefront of complex dynamics. Their physical consistency depends on the constraints built into the representation, the reference data, and the deployment. The following criteria apply to both broadly pretrained and system-specific models.

Low test errors and stable short trajectories are useful checks, but they are insufficient on their own. Validation should connect architectural properties, numerical behavior, and the observables supporting the scientific claim. Three recurring issues are size consistency, nonlocal interactions, and the relationship between energies and derivatives.

### 5.1. Size consistency and physical completeness

Size consistency requires that the energy of two noninteracting subsystems equals the sum of their separate energies. For a periodically repeated system evaluated with consistent boundary conditions, an extensive short-range energy model should scale proportionally with the number of repeated cells; finite-size effects and long-range electrostatics can complicate comparisons between differently sized cells.

For local MPNNs and equivariant potentials of the form $E = \sum_i E_i$, additivity is built into the architecture: two noninteracting replicas with identical local environments have exactly additive predicted energies, apart from numerical and neighbor-list effects. This architectural size consistency does not imply physical completeness. If charge transfer, polarization, or electrostatics depends on the global cell, a finite-cutoff model can predict an incorrect thermodynamic-limit energy even while remaining mathematically extensive.

The relevant distinction is therefore between architectural extensivity and correct treatment of nonlocal physics. Charge-equilibration, Ewald-type, reciprocal-space, or self-consistent-field components can be necessary when observables depend on global charge redistribution. They supplement rather than create the basic size consistency of a local energy sum, and their need should be established for the target system and observable.

### 5.2. Incorporating long-range interactions

The success of local MLIPs is related to Walter Kohn's nearsightedness principle for electronic matter. SOAP descriptors and many MACE implementations use a finite local cutoff, often of a few ångströms, although the effective receptive field and chosen cutoff are model- and application-dependent.

For ionic liquids, polar surfaces, ferroelectrics, and biomolecules, long-range electrostatics and dispersion may materially affect the target observables. Coulomb interactions usually require a controlled long-range treatment under periodic boundary conditions; dispersion decays faster and can sometimes be truncated with switching or tail corrections. A strictly local graph cannot represent a global polarization field that is not encoded in its inputs, which matters for dielectric response and non-analytic phonon effects such as LO–TO splitting. Methods that incorporate nonlocal information include LODE-type features [203], Ewald message passing [204, 205], reciprocal-space networks [206], charge equilibration [207], and Laplacian-eigenbasis electrostatics [208]. Multilayer ACE [209] instead extends semilocal or intermediate-range interactions; expanding a finite receptive field is distinct from specifying global electrostatics and its boundary conditions. NequIP-LR combines equivariant local features with global charge redistribution [210]. These methods trade additional assumptions and computational cost for a broader physical receptive field. Their formulation and boundary conditions determine which nonlocal interactions are represented.

Physical scaling priors can constrain local representations, as discussed in Section 2.3. Such constraints complement, but do not replace, a treatment of nonlocal interactions when these affect the target observable.

### 5.3. Energy–force consistency and response properties

For a time-independent Hamiltonian description in the microcanonical ensemble (NVE), forces should derive from the modeled energy. For a continuously differentiable force field, symmetry of its Jacobian, $\frac{\partial F_a}{\partial R_b} = \frac{\partial F_b}{\partial R_a}$ for all Cartesian coordinates a and b, is the local integrability condition in the 3N-dimensional configuration space. A global potential additionally depends on the domain; on a simply connected domain this symmetry is sufficient. A suitable integrator and timestep are also required to control numerical drift. Thermostatted, driven, or stochastic dynamics impose different requirements and should not be assessed by NVE energy conservation alone.

Unconstrained direct-force regression does not guarantee that the learned vector field is conservative, so closed-loop work and NVE energy drift should be tested [28, 111]. Direct force prediction is not intrinsically nonconservative: conservative parameterizations or integrability constraints can be imposed. The common energy-based construction $\mathbf{F} = -\nabla E$, used by models such as MACE and NequIP, guarantees conservativity for a differentiable, time-independent energy model up to numerical errors.

Energy and force errors alone do not characterize observables that depend on curvature, stress, or activated pathways. Phonon spectra, elastic response, and transition-state curvatures require appropriate derivative tests [102, 124]. Force- or gradient-aware training can improve learning [211], while Hessian information is useful when justified by the target observable.

For metals, this principle leads to concrete mechanistic tests. Zhang et al. benchmarked MLIPs for ferromagnetic iron using screw-dislocation and cleavage properties alongside conventional errors [212]. Dislocation-core structure, migration barriers, and crack response probed distinctions that aggregate metrics could miss. Their zero-temperature, quasi-static fracture calculations complement finite-temperature MD by isolating defect energetics and crack-tip response.

## 6. Outlook

Future progress will depend on matching representational capacity, physical state variables, reference quality, sampling, and computational cost to the intended application. The most useful advances will connect these choices to measurable improvements in validated observables and accessible simulation scales.

Geometry and atomic species do not uniquely specify every state relevant to materials dynamics. Charge, spin, electrode potential, or external fields may be needed when distinct electronic conditions share similar structures. Some current models already include these variables. Future work should establish when such extensions are necessary and how to maintain consistent energies, forces, and responses across the resulting state space.

Architectural encoding of physical scaling laws can complement explicit state variables. A useful research direction is to test whether similarly economical representations can incorporate polarization or magnetic interactions while retaining the required boundary-condition dependence. Any gains in parameter efficiency or transferability should be demonstrated against matched baselines and the relevant dynamical observables.

Beyond atomistic observables, MLIPs can supply defect energetics, mobilities, and constitutive information to phase-field, continuum, or finite-element models. Metallic deformation and fracture studies offer concrete routes to this scale bridging. The transferred quantities inherit reference, fit, sampling, and finite-size errors, as well as coarse-graining assumptions. Propagating those uncertainties is essential when atomistic mechanisms are used to predict larger-scale materials behavior.

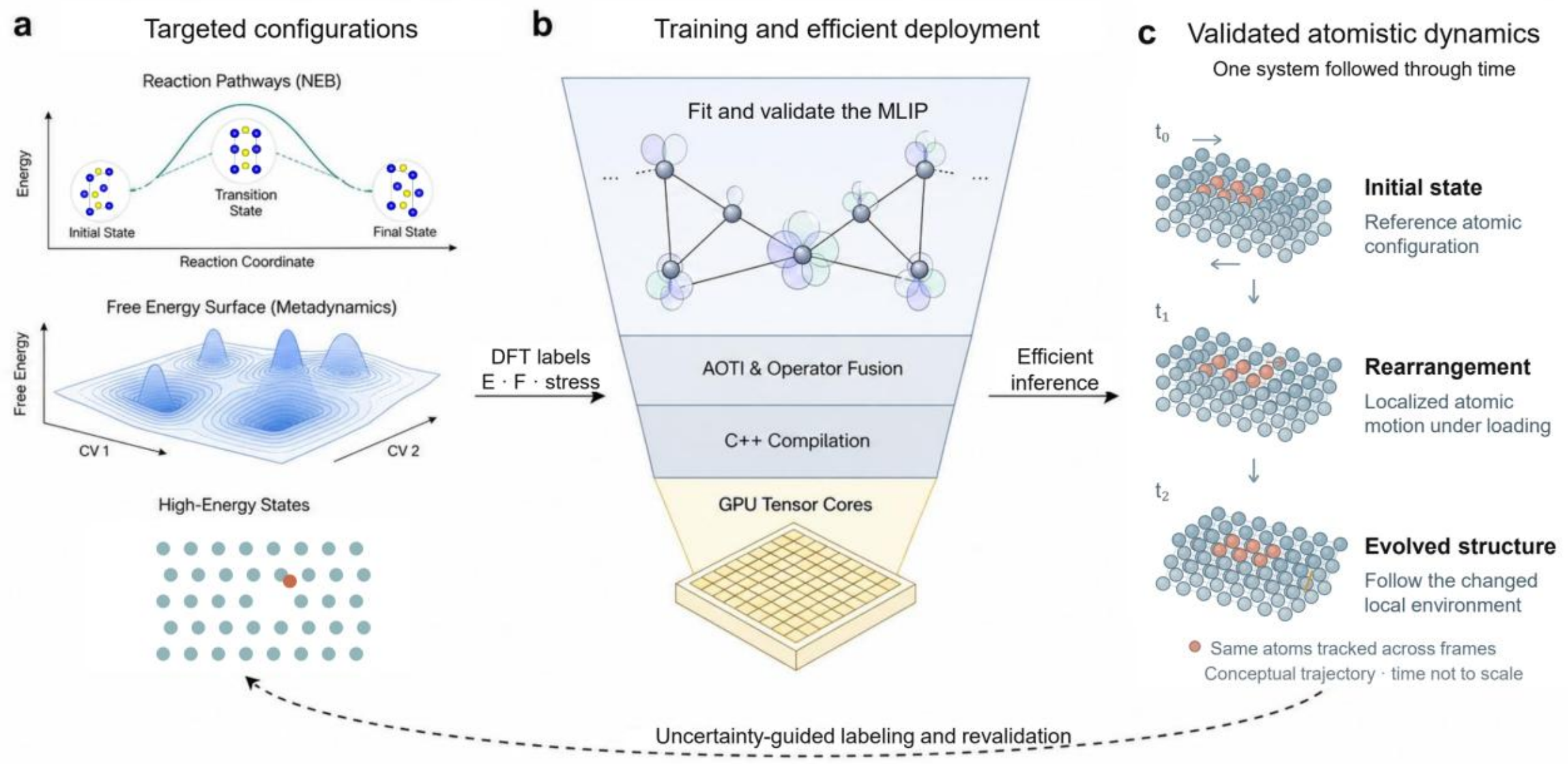

Figure 6. Integrated closed-loop paradigm for next-generation MLIPs. (a) Physics-driven sampling strategies. Path-search methods such as the nudged elastic band (NEB) resolve selected transition pathways, while enhanced sampling such as metadynamics explores free-energy landscapes. These methods can improve coverage of relevant configurations, including defect transitions. Reference-label fidelity depends separately on the electronic-structure method and numerical convergence. (b) Hardware-aware compilation and software-hardware co-design. Ahead-of-time inference (AOTI), operator fusion, and specialized kernels can reduce evaluation overhead and improve hardware utilization. Use of GPU Tensor Cores and the resulting gains depend on the operators, precision, and implementation; compiled models should retain the intended symmetry and conservation properties. (c) Validated atomistic dynamics. Three schematic configurations follow the same atomic system from its initial state through local rearrangement to an evolved structure. Coral-colored atoms mark the same atoms across frames; the configurations and time labels are illustrative rather than a recorded trajectory. The dashed arrow illustrates a proposed uncertainty-driven active-learning loop in which simulations flag candidate extrapolative environments for targeted labeling and revalidation. The accessible system size and duration remain conditional on model fidelity, sampling, and implementation.

The roadmap also depends on a robust engineering ecosystem. PyTorch and JAX accelerate model development, but interpreter overhead, kernel-launch cost, memory movement, and distributed communication can become important for small local workloads or very large simulations. Their relative importance depends on the compiled execution path and hardware.

Software–hardware co-design can improve throughput through ahead-of-time compilation, operator fusion, and communication-aware partitioning (Figure 6b). Exported C++ or compiled runtimes remove Python from the inner evaluation loop. cuEquivariance optimizes equivariant tensor operations [213], while DP Compress accelerates compatible Deep Potential models through descriptor tabulation and kernel fusion [214]. FastCHGNet, FlashTP, and NequIP 2.0 pursue related improvements in model evaluation [215, 216, 217]. Their practical gains depend on hardware, interaction range, precision, and workload. Larger simulations will benefit scientific interpretation when these gains are coupled to validated potentials and adequate sampling.

The diversity of current MLIPs creates opportunities to select models by the physical and computational demands of an application, rather than by architectural generation alone.

Physics-driven specialization extends the practical reach of MLIPs by aligning model design, reference data, and deployment with the demands of materials dynamics. Broad coverage establishes a useful starting point, while reliable application requires suitable physical variables, reference fidelity, configurational coverage, conservation properties, and evaluation cost. A pretrained model that meets these requirements can be used directly. Where a gap remains, targeted adaptation should address it and demonstrate a benefit through the intended observables and suitable task-compatible comparisons. The four application contexts in Section 4 and the validation criteria in Section 5 provide a framework for making those decisions.

A useful next step is to make those requirements explicit. Reporting the validated configurations, observables, boundary conditions, and computational settings will help connect model benchmarks to scientific claims. This would support a more reliable progression from electronic-structure reference data to atomistic mechanisms and, where justified, larger-scale predictions.

## Competing interests

The authors declare no competing interests.

## References

[1] Kohn, W. & Sham, L. J. Self-consistent equations including exchange and correlation effects. *Physical Review* **140**, A1133 (1965).

[2] Car, R. & Parrinello, M. Unified approach for molecular dynamics and density-functional theory. *Physical Review Letters* **55**, 2471–2474 (1985).

[3] Goedecker, S. Linear scaling electronic structure methods. *Reviews of Modern Physics* **71**, 1085–1123 (1999). https://doi.org/10.1103/RevModPhys.71.1085

[4] Marx, D. & Hutter, J. *Ab Initio Molecular Dynamics: Basic Theory and Advanced Methods*. (Cambridge University Press, Cambridge, U.K., 2009).

[5] Jones, J. E. On the determination of molecular fields. *Proceedings of the Royal Society of London. Series A* **106**, 463–477 (1924).

[6] Born, M. & Mayer, J. E. Zur gittertheorie der ionenkristalle. *Zeitschrift für Physik* **75**, 1–18 (1932).

[7] Buckingham, R. A. The classical equation of state of gaseous helium, neon and argon. *Proceedings of the Royal Society of London. Series A* **168**, 264–283 (1938).

[8] Stillinger, F. H. & Weber, T. A. Computer simulation of local order in condensed phases of silicon. *Physical Review B* **31**, 5262 (1985).

[9] Tersoff, J. New empirical approach for the structure and energy of covalent systems. *Physical Review B* **37**, 6991 (1988).

[10] Daw, M. S. & Baskes, M. I. Embedded-atom method: Derivation and application to impurities, surfaces, and other defects in metals. *Physical Review B* **29**, 6443 (1984).

[11] Duin, A. C. T. van, Dasgupta, S., Lorant, F. & Goddard, W. A. ReaxFF: A reactive force field for hydrocarbons. *Journal of Physical Chemistry A* **105**, 9396–9409 (2001).

[12] Yu, J., Sinnott, S. B. & Phillpot, S. R. Charge optimized many-body potential for the Si/SiO2 system. *Physical Review B* **75**, 085311 (2007).

[13] Liang, T., Devine, B., Phillpot, S. R. & Sinnott, S. B. Variable charge reactive potential for hydrocarbons to simulate organic-copper interactions. *The Journal of Physical Chemistry A* **116**, 7976–7991 (2012).

[14] Kim, J., Cho, H., Jeon, H., Jung, J. & Han, S. Reactive machine learning interatomic potentials for chemistry and materials science. *Chemical Reviews* **126**, 4467–4510 (2026).

[15] Behler, J. Perspective: Machine learning potentials for atomistic simulations. *The Journal of Chemical Physics* **145**, 170901 (2016).

[16] Behler, J. & Parrinello, M. Generalized neural-network representation of high-dimensional potential-energy surfaces. *Physical Review Letters* **98**, 146401 (2007).

[17] Bartók, A. P., Payne, M. C., Kondor, R. & Csányi, G. Gaussian approximation potentials: The accuracy of quantum mechanics, without the electrons. *Physical Review Letters* **104**, 136403 (2010).

[18] Deringer, V. L., Caro, M. A. & Csányi, G. Machine learning interatomic potentials as emerging tools for materials science. *Advanced Materials* **31**, 1902765 (2019).

[19] Jia, W. *et al.* Pushing the limit of molecular dynamics with ab initio accuracy to 100 million atoms with machine learning. in *SC20: International conference for high performance computing, networking, storage and analysis* (IEEE, 2020).

[20] Chen, C. & Ong, S. P. A universal graph deep learning interatomic potential for the periodic table. *Nature Computational Science* **2**, 718–728 (2022).

[21] Deng, B. *et al.* CHGNet as a pretrained universal neural network potential for charge-informed atomistic modelling. *Nature Machine Intelligence* **5**, 1031–1041 (2023).

[22] Batatia, I. *et al.* A foundation model for atomistic materials chemistry. *The Journal of Chemical Physics* **163**, 184110 (2025).

[23] Jain, A. *et al.* Commentary: The materials project: A materials genome approach to accelerating materials innovation. *APL Materials* **1**, 011002 (2013).

[24] Saal, J. E., Kirklin, S., Aykol, M., Meredig, B. & Wolverton, C. Materials design and discovery with high-throughput density functional theory: The open quantum materials database (OQMD). *JOM* **65**, 1501–1509 (2013).

[25] Kirklin, S. *et al.* The open quantum materials database (OQMD): Assessing the accuracy of DFT formation energies. *npj Computational Materials* **1**, 15010 (2015).

[26] Lal, A., Breebaart, R. S., Bolhuis, P. G. & Meijer, E. J. Discovering reaction mechanisms with transition path sampling-based active learning of machine-learned potentials. *arXiv preprint* arXiv:2605.03737 (2026).

[27] Fu, X. *et al.* Forces are not enough: Benchmark and critical evaluation for machine learning force fields with molecular simulations. *Transactions on Machine Learning Research* (2023).

[28] Morrow, J. D., Gardner, J. L. A. & Deringer, V. L. How to validate machine-learned interatomic potentials. *The Journal of Chemical Physics* **158**, 121501 (2023). https://doi.org/10.1063/5.0139611

[29] Schmitz, N. F., Müller, K.-R. & Chmiela, S. Algorithmic differentiation for automated modeling of machine learned force fields. *The Journal of Physical Chemistry Letters* **13**, 10020–10028 (2022).

[30] Amin, I., Raja, S. & Krishnapriyan, A. S. Towards fast, specialized machine learning force fields: Distilling foundation models via energy hessians. in *The thirteenth international conference on learning representations (ICLR)* (2025).

[31] Liu, X., Zeng, K., Wang, Y. & Zhao, T. A study on the fine-tuning performance of universal machine-learned interatomic potentials (u-MLIPs). *arXiv preprint arXiv:2506.07401* https://doi.org/10.48550/arXiv.2506.07401 (2025) doi:10.48550/arXiv.2506.07401.

[32] Bartók, A. P., Kondor, R. & Csányi, G. On representing chemical environments. *Physical Review B* **87**, 184115 (2013).

[33] Deringer, V. L. *et al.* Gaussian process regression for materials and molecules. *Chemical Reviews* **121**, 10073–10141 (2021).

[34] Klawohn, S. *et al.* Gaussian approximation potentials: Theory, software implementation and application examples. *The Journal of Chemical Physics* **159**, 171502 (2023).

[35] Ulanov, E., Qadir, G. A., Riedmiller, K., Friederich, P. & Gräter, F. Predicting hydrogen atom transfer energy barriers using gaussian process regression. *Digital Discovery* **4**, 513–522 (2025).

[36] Jinnouchi, R., Karsai, F. & Kresse, G. On-the-fly machine learning force field generation: Application to melting points. *Physical Review B* **100**, 014105 (2019).

[37] Caro, M. A., Csányi, G., Laurila, T. & Deringer, V. L. Machine learning driven simulated deposition of carbon films: From low-density to diamond-like amorphous carbon. *Physical Review B* **102**, 174201 (2020).

[38] Szlachta, W. J., Bartók, A. P. & Csányi, G. Accuracy and transferability of gaussian approximation potential models for tungsten. *Physical Review B* **90**, 104108 (2014).

[39] Bartók, A. P., Kermode, J., Bernstein, N. & Csányi, G. Machine learning a general-purpose interatomic potential for silicon. *Physical Review X* **8**, 041048 (2018).

[40] Deringer, V. L. & Csányi, G. Machine learning based interatomic potential for amorphous carbon. *Physical Review B* **95**, 094203 (2017).

[41] Chmiela, S. *et al.* Machine learning of accurate energy-conserving molecular force fields. *Science Advances* **3**, e1603015 (2017).

[42] Chmiela, S., Sauceda, H. E., Müller, K.-R. & Tkatchenko, A. Towards exact molecular dynamics simulations with machine-learned force fields. *Nature Communications* **9**, 3887 (2018). https://doi.org/10.1038/s41467-018-06169-2

[43] Thompson, A. P., Swiler, L. P., Trott, C. R., Foiles, S. M. & Tucker, G. J. Spectral neighbor analysis method for automated generation of quantum-accurate interatomic potentials. *Journal of Computational Physics* **285**, 316–330 (2015).

[44] Wood, M. A. & Thompson, A. P. Extending the accuracy of the SNAP interatomic potential form. *The Journal of Chemical Physics* **148**, 241721 (2018).

[45] Cusentino, M. A., Wood, M. A. & Thompson, A. P. Explicit Multielement Extension of the Spectral Neighbor Analysis Potential for Chemically Complex Systems. *The Journal of Physical Chemistry A* **124**, 5456–5464 (2020). https://doi.org/10.1021/acs.jpca.0c02450

[46] Shapeev, A. V. Moment tensor potentials: A class of systematically improvable interatomic potentials. *Multiscale Modeling & Simulation* **14**, 1153–1173 (2016).

[47] Podryabinkin, E. V. & Shapeev, A. V. Active learning of linearly parametrized interatomic potentials. *Computational Materials Science* **140**, 171–180 (2017).

[48] Novikov, I. S., Gubaev, K., Podryabinkin, E. V. & Shapeev, A. V. The MLIP package: Moment tensor potentials with MPI and active learning. *Machine Learning: Science and Technology* **2**, 025002 (2021).

[49] Gubaev, K., Podryabinkin, E. V., Hart, G. L. & Shapeev, A. V. Accelerating high-throughput searches for new alloys with active learning of interatomic potentials. *Computational Materials Science* **156**, 148–156 (2019).

[50] Novikov, I. S., Shapeev, A. V. & Suleimanov, Y. V. Ring polymer molecular dynamics and active learning of moment tensor potential for gas-phase barrierless reactions: Application to s + H2. *The Journal of Chemical Physics* **151**, 224105 (2019).

[51] Chen, B. W. J., Wang, B., Sullivan, M. B., Borgna, A. & Zhang, J. Unraveling the synergistic effect of Re and Cs promoters on ethylene epoxidation over silver catalysts with machine learning-accelerated first-principles simulations. *ACS Catalysis* **12**, 2540–2551 (2022).

[52] Drautz, R. Atomic cluster expansion for accurate and transferable interatomic potentials. *Physical Review B* **99**, 014104 (2019).

[53] Seko, A., Togo, A. & Tanaka, I. Group-theoretical high-order rotational invariants for structural representations: Application to linearized machine learning interatomic potential. *Physical Review B* **99**, 214108 (2019).

[54] Dusson, G. *et al.* Atomic cluster expansion: Completeness, efficiency and stability. *Journal of Computational Physics* **454**, 110946 (2022).

[55] Lysogorskiy, Y. *et al.* Performant implementation of the atomic cluster expansion (PACE) and application to copper and silicon. *npj Computational Materials* **7**, 97 (2021).

[56] Kovács, D. P. *et al.* Linear atomic cluster expansion force fields for organic molecules: Beyond RMSE. *Journal of Chemical Theory and Computation* **17**, 7696–7711 (2021).

[57] Willman, J. T., Perriot, R. & Ticknor, C. Accurate and efficient parameterization of an atomic cluster expansion (ACE) potential for ammonia under extreme conditions. *The Journal of Chemical Physics* **162**, 144316 (2025).

[58] Zhang, H., Juraskova, V. & Duarte, F. Modelling chemical processes in explicit solvents with machine learning potentials. *Nature Communications* **15**, 6114 (2024).

[59] Blank, T. B., Brown, S. D., Calhoun, A. W. & Doren, D. J. Neural network models of potential energy surfaces. *The Journal of Chemical Physics* **103**, 4129–4137 (1995).

[60] Behler, J. Atom-centered symmetry functions for constructing high-dimensional neural network potentials. *The Journal of Chemical Physics* **134**, 074106 (2011).

[61] Romano, S. *et al.* Structure and dynamics of the magnetite(001)/water interface from molecular dynamics simulations based on a neural network potential. *Journal of Chemical Theory and Computation* **21**, 1951–1960 (2025).

[62] Ghasemi, S. A., Hofstetter, A., Saha, S. & Goedecker, S. Interatomic potentials for ionic systems with density functional accuracy based on charge densities obtained by a neural network. *Physical Review B* **92**, 045131 (2015).

[63] Smith, J. S., Isayev, O. & Roitberg, A. E. ANI-1: An extensible neural network potential with DFT accuracy at force field computational cost. *Chemical Science* **8**, 3192–3203 (2017).

[64] Smith, J. S. *et al.* Approaching coupled cluster accuracy with a general-purpose neural network potential through transfer learning. *Nature Communications* **10**, 2903 (2019).

[65] Devereux, C. *et al.* Extending the applicability of the ANI deep learning molecular potential to sulfur and halogens. *Journal of Chemical Theory and Computation* **16**, 4192–4202 (2020).

[66] Gao, X., Ramezanghorbani, F., Isayev, O., Smith, J. S. & Roitberg, A. E. TorchANI: A free and open source PyTorch-based deep learning implementation of the ANI neural network potentials. *Journal of Chemical Information and Modeling* **60**, 3408–3415 (2020).

[67] Anstine, D. M., Zubatyuk, R. & Isayev, O. AIMNet2: A neural network potential to meet your neutral, charged, organic, and elemental-organic needs. *Chemical Science* **16**, 10228–10244 (2025).

[68] Fan, Z. et al. Neuroevolution machine learning potentials: Combining high accuracy and low cost in atomistic simulations and application to heat transport. *Physical Review B* **104**, 104309 (2021). https://doi.org/10.1103/PhysRevB.104.104309

[69] Fan, Z. et al. GPUMD: A package for constructing accurate machine-learned potentials and performing highly efficient atomistic simulations. *The Journal of Chemical Physics* **157**, 114801 (2022). https://doi.org/10.1063/5.0106617

[70] Zhang, L., Han, J., Wang, H., Car, R. & E, W. Deep potential molecular dynamics: A scalable model with the accuracy of quantum mechanics. *Physical Review Letters* **120**, 143001 (2018).

[71] Zhang, L., Han, J., Wang, H., Saidi, W. A. & Car, R. End-to-end symmetry preserving inter-atomic potential energy model for finite and extended systems. in *Advances in neural information processing systems* vol. 31 4436–4446 (2018).

[72] Wang, H., Zhang, L., Han, J. & E, W. DeePMD-kit: A deep learning package for many-body potential energy representation and molecular dynamics. *Computer Physics Communications* **228**, 178–184 (2018).

[73] Zuo, Y. *et al.* Performance and cost assessment of machine learning interatomic potentials. *The Journal of Physical Chemistry A* **124**, 731–745 (2020).

[74] Gilmer, J., Schoenholz, S. S., Riley, P. F., Vinyals, O. & Dahl, G. E. Neural message passing for quantum chemistry. in *International conference on machine learning (ICML)* vol. 70 1263–1272 (PMLR, 2017).

[75] Kang, S. How graph neural network interatomic potentials extrapolate: Role of the message-passing algorithm. *The Journal of Chemical Physics* **161**, 244102 (2024).

[76] Schütt, K. T. *et al.* SchNet: A continuous-filter convolutional neural network for modeling quantum interactions. in *Advances in neural information processing systems (NeurIPS)* vol. 30 (2017).

[77] Grimme, S., Antony, J., Ehrlich, S. & Krieg, H. A consistent and accurate ab initio parametrization of density functional dispersion correction (DFT-d) for the 94 elements h-pu. *The Journal of Chemical Physics* **132**, 154104 (2010).

[78] Unke, O. T. & Meuwly, M. PhysNet: A neural network for predicting energies, forces, dipole moments, and partial charges. *Journal of Chemical Theory and Computation* **15**, 3678–3693 (2019).

[79] Rivero, U., Unke, O. T., Meuwly, M. & Willitsch, S. Reactive atomistic simulations of Diels–Alder reactions: The importance of molecular rotations. *The Journal of Chemical Physics* **151**, 104301 (2019).

[80] Westermayr, J., Gastegger, M. & Marquetand, P. Combining SchNet and SHARC: The SchNarc machine learning approach for excited-state dynamics. *The Journal of Physical Chemistry Letters* **11**, 3828–3834 (2020).

[81] Gastegger, M., Schütt, K. T. & Müller, K.-R. Machine learning of solvent effects on molecular spectra and reactions. *Chemical Science* **12**, 11473–11483 (2021).

[82] Klicpera, J., Groß, J. & Günnemann, S. Directional message passing for molecular graphs. in *International conference on learning representations (ICLR)* (2020).

[83] Gasteiger, J., Giri, S., Margraf, J. T. & Günnemann, S. Fast and uncertainty-aware directional message passing for non-equilibrium molecules. in *Machine learning for molecules workshop at NeurIPS* (2020).

[84] Liu, Y. *et al.* Spherical message passing for 3D molecular graphs. in *International conference on learning representations (ICLR)* (2022).

[85] Tawfik, S. A. *et al.* Embedding material graphs using the electron-ion potential: Application to material fracture. *arXiv preprint arXiv:2402.10931* https://doi.org/10.48550/arXiv.2402.10931 (2024) doi:10.48550/arXiv.2402.10931.

[86] Thomas, N. *et al.* Tensor field networks: Rotation- and translation-equivariant neural networks for 3D point clouds. *arXiv preprint arXiv:1802.08219* (2018).

[87] Anderson, B., Hy, T.-S. & Kondor, R. Cormorant: Covariant molecular neural networks. in *Advances in neural information processing systems (NeurIPS)* vol. 32 14537–14546 (2019).

[88] Batzner, S. *et al.* E(3)-equivariant graph neural networks for data-efficient and accurate interatomic potentials. *Nature Communications* **13**, 2453 (2022).

[89] Schütt, K., Unke, O. & Gastegger, M. Equivariant message passing for the prediction of tensorial properties and molecular spectra. in *International conference on machine learning (ICML)* 9377–9388 (PMLR, 2021).

[90] Haghighatlari, M. *et al.* NewtonNet: A newtonian message passing network for deep learning of interatomic potentials and forces. *Digital Discovery* **1**, 333–343 (2022).

[91] Batatia, I., Kovács, D. P., Simm, G. N. C., Ortner, C. & Csányi, G. MACE: Higher order equivariant message passing neural networks for fast and accurate force fields. in *Advances in neural information processing systems (NeurIPS)* vol. 35 11423–11436 (2022).

[92] Musaelian, A. *et al.* Learning local equivariant representations for large-scale atomistic dynamics. *Nature Communications* **14**, 579 (2023).

[93] Hu, Y. et al. Global universal scaling and ultrasmall parameterization in machine-learning interatomic potentials with superlinearity. *Proceedings of the National Academy of Sciences* **122**, e2503439122 (2025). https://doi.org/10.1073/pnas.2503439122

[94] Wang, J. *et al.* E(n)-equivariant cartesian tensor message passing interatomic potential. *Nature Communications* **15**, 7607 (2024).

[95] Xu, Z., Xie, W. & Hu, P. Spectral/spatial tensor atomic cluster expansion with universal embeddings in cartesian space. *arXiv preprint arXiv:2509.14961* https://doi.org/10.48550/arXiv.2509.14961 (2025) doi:10.48550/arXiv.2509.14961.

[96] Simeon, G. & De Fabritiis, G. TensorNet: Cartesian tensor representations for efficient learning of molecular potentials. in *Advances in neural information processing systems (NeurIPS)* vol. 36 68112–68128 (2023).

[97] Cheng, B. Cartesian atomic cluster expansion for machine learning interatomic potentials. *npj Computational Materials* **10**, 157 (2024). https://doi.org/10.1038/s41524-024-01332-4

[98] Wen, M., Huang, W.-F., Dai, J. & Adhikari, S. Cartesian atomic moment machine learning interatomic potentials. *npj Computational Materials* **11**, 128 (2025).

[99] Liao, Y.-L., Wood, B., Das, A. & Smidt, T. EquiformerV2: Improved equivariant transformer for scaling to higher-degree representations. in *International conference on learning representations (ICLR)* (2024).

[100] Passaro, S. & Zitnick, C. L. Reducing SO(3) convolutions to SO(2) for efficient equivariant GNNs. in *Proceedings of the 40th international conference on machine learning (ICML)* 27420–27438 (2023).

[101] Frank, J. T., Unke, O. T., Müller, K.-R. & Chmiela, S. A euclidean transformer for fast and stable machine learned force fields. *Nature Communications* **15**, 6539 (2024).

[102] Fu, X. *et al.* Learning smooth and expressive interatomic potentials for physical property prediction. *arXiv preprint arXiv:2502.12147* (2025).

[103] Qu, E. & Krishnapriyan, A. S. The importance of being scalable: Improving the speed and accuracy of neural network interatomic potentials across chemical domains. in *Advances in neural information processing systems* vol. 37 139030–139053 (2024).

[104] Brehmer, J., Behrends, S., Haan, P. de & Cohen, T. Does equivariance matter at scale? *arXiv preprint arXiv:2410.23179* (2025).

[105] Li, H., Wang, Z., Zou, N., Zhao, M. & Xu, S. Deep-learning density functional theory hamiltonian for efficient ab initio electronic-structure calculation. *Nature Computational Science* **2**, 367–377 (2022).

[106] Eckhoff, M. & Behler, J. High-dimensional neural network potentials for magnetic systems using spin-dependent atom-centered symmetry functions. *npj Computational Materials* **7**, 170 (2021).

[107] Ko, T. W., Finkler, J. A., Goedecker, S. & Behler, J. A fourth-generation high-dimensional neural network potential with accurate electrostatics including non-local charge transfer. *Nature Communications* **12**, 398 (2021).

[108] Yang, T. *et al.* Deep learning illuminates spin and lattice interaction in magnetic materials. *Physical Review B* **110**, 064427 (2024).

[109] Kotykhov, A. S. *et al.* Actively trained magnetic moment tensor potentials for mechanical, dynamical, and thermal properties of paramagnetic CrN. *Physical Review B* **111**, 094438 (2025).

[110] Rhodes, B. et al. Orb-v3: atomistic simulation at scale. arXiv preprint arXiv:2504.06231 (2025). https://doi.org/10.48550/arXiv.2504.06231

[111] Bigi, F., Langer, M. & Ceriotti, M. The dark side of the forces: Assessing non-conservative force models for atomistic machine learning. in *Proceedings of the 42nd international conference on machine learning* 4384–4414 (2025).

[112] Barros-Luque, L. et al. The Open Materials 2024 (OMat24) inorganic materials dataset and models. *Nature Computational Science* **6**, 642–652 (2026). https://doi.org/10.1038/s43588-026-00996-w

[113] Yang, H. et al. MatterSim: A deep learning atomistic model across elements, temperatures and pressures. Preprint at https://arxiv.org/abs/2405.04967 (2024).

[114] Deng, B. *et al.* Systematic softening in universal machine learning interatomic potentials. *npj Computational Materials* **11**, 9 (2025).

[115] Focassio, B., Freitas, L. P. M. & Schleder, G. R. Performance assessment of universal machine learning interatomic potentials: Challenges and directions for materials' surfaces. *ACS Applied Materials & Interfaces* **17**, 13111–13121 (2025).

[116] Yoo, D., Jung, J., Jeong, W. & Han, S. Metadynamics sampling in atomic environment space for collecting training data for machine learning potentials. *npj Computational Materials* **7**, 131 (2021).

[117] Waters, M. J. & Rondinelli, J. M. Benchmarking structural evolution methods for training of machine learned interatomic potentials. *Journal of Physics: Condensed Matter* **34**, 385901 (2022).

[118] Henkelman, G. & Jónsson, H. Improved tangent estimate in the nudged elastic band method for finding minimum energy paths and saddle points. *The Journal of Chemical Physics* **113**, 9978–9985 (2000).

[119] Henkelman, G., Uberuaga, B. P. & Jónsson, H. A climbing image nudged elastic band method for finding saddle points and minimum energy paths. *The Journal of Chemical Physics* **113**, 9901–9904 (2000).

[120] Wander, B., Shuaibi, M., Kitchin, J. R., Ulissi, Z. W. & Zitnick, C. L. CatTSunami: Accelerating transition state energy calculations with pretrained graph neural networks. *ACS Catalysis* **15**, 5283–5294 (2025).

[121] Schreiner, M., Bhowmik, A., Vegge, T., Busk, J. & Winther, O. Transition1x - a dataset for building generalizable reactive machine learning potentials. *Scientific Data* **9**, 779 (2022).

[122] Schreiner, M., Bhowmik, A., Vegge, T., Jørgensen, P. B. & Winther, O. NeuralNEB—neural networks can find reaction paths fast. *Machine Learning: Science and Technology* **3**, 045022 (2022).

[123] Barrett, R. & Westermayr, J. Reinforcement learning for traversing chemical structure space: Optimizing transition states and minimum energy paths of molecules. *The Journal of Physical Chemistry Letters* **15**, 349–356 (2024).

[124] Rodriguez, A., Smith, J. S. & Mendoza-Cortes, J. L. Does hessian data improve the performance of machine learning potentials? *Journal of Chemical Theory and Computation* **21**, 6698–6710 (2025).

[125] Smith, J. S. *et al.* The ANI-1ccx and ANI-1x data sets, coupled-cluster and density functional theory properties for molecules. *Scientific Data* **7**, 134 (2020).

[126] Hoja, J. *et al.* QM7-x, a comprehensive dataset of quantum-mechanical properties spanning the chemical space of small organic molecules. *Scientific Data* **8**, 43 (2021).

[127] Tang, K. *et al.* Accelerating transition state search and ligand screening for organometallic catalysis with reactive machine learning potential. *Journal of Chemical Theory and Computation* **21**, 8933–8949 (2025).

[128] Henkelman, G. & Jónsson, H. A dimer method for finding saddle points on high dimensional potential surfaces using only first derivatives. *The Journal of Chemical Physics* **111**, 7010–7022 (1999).

[129] Jay, A. *et al.* Finding reaction pathways and transition states: R-ARTn and d-ARTn as an efficient and versatile alternative to string approaches. *Journal of Chemical Theory and Computation* **16**, 6726–6734 (2020).

[130] Shang, C. & Liu, Z.-P. Stochastic surface walking method for structure prediction and pathway searching. *Journal of Chemical Theory and Computation* **9**, 1838–1845 (2013).

[131] Chen, D., Kang, P.-L. & Liu, Z.-P. Active site of catalytic ethene epoxidation: Machine-learning global pathway sampling rules out the metal sites. *ACS Catalysis* **11**, 8317–8326 (2021).

[132] Xie, X.-T. *et al.* LASP to the future of atomic simulation: Intelligence and automation. *Precision Chemistry* **2**, 612–627 (2024).

[133] Ryckaert, J.-P., Ciccotti, G. & Berendsen, H. J. C. Numerical integration of the cartesian equations of motion of a system with constraints: Molecular dynamics of n-alkanes. *Journal of Computational Physics* **23**, 327–341 (1977).

[134] Zhang, S., Zubatyuk, R., Yang, Y., Roitberg, A. & Isayev, O. ANI-1xBB: An ANI-based reactive potential for small organic molecules. *Journal of Chemical Theory and Computation* **21**, 4365–4374 (2025).

[135] Hong, C. *et al.* Atomistic simulation of HF etching process of amorphous Si3N4 using machine learning potential. *ACS Applied Materials & Interfaces* **16**, 48457–48469 (2024).

[136] Torrie, G. M. & Valleau, J. P. Nonphysical sampling distributions in monte carlo free-energy estimation: Umbrella sampling. *Journal of Computational Physics* **23**, 187–199 (1977).

[137] Nakata, H., Filatov Gulak, M. & Choi, C. H. Accelerated deep learning dynamics for atomic layer deposition of al(me)3 and water on OH/si(111). *ACS Applied Materials & Interfaces* **14**, 26116–26127 (2022).

[138] Stocker, S. *et al.* Estimating free energy barriers for heterogeneous catalytic reactions with machine learning potentials and umbrella integration. *Journal of Chemical Theory and Computation* **19**, 6796–6804 (2023).

[139] Laio, A. & Parrinello, M. Escaping free-energy minima. *Proceedings of the National Academy of Sciences* **99**, 12562–12566 (2002).

[140] Pracht, P., Bohle, F. & Grimme, S. Automated exploration of the low-energy chemical space with fast quantum chemical methods. *Physical Chemistry Chemical Physics* **22**, 7169–7192 (2020).

[141] David, R. *et al.* ArcaNN: Automated enhanced sampling generation of training sets for chemically reactive machine learning interatomic potentials. *Digital Discovery* **4**, 54–72 (2025).

[142] Miwa, K. Restricted multicanonical sampling for machine learning potential construction. *Physical Review B* **107**, 054313 (2023).

[143] Zhao, Q. & Savoie, B. M. Simultaneously improving reaction coverage and computational cost in automated reaction prediction tasks. *Nature Computational Science* **1**, 479–490 (2021).

[144] Zhao, Q. *et al.* Comprehensive exploration of graphically defined reaction spaces. *Scientific Data* **10**, 145 (2023).

[145] Zhang, Y. *et al.* DP-GEN: A concurrent learning platform for the generation of reliable deep learning based potential energy models. *Computer Physics Communications* **253**, 107206 (2020).

[146] Zhu, A., Batzner, S., Musaelian, A. & Kozinsky, B. Fast uncertainty estimates in deep learning interatomic potentials. *The Journal of Chemical Physics* **158**, 164111 (2023).

[147] Kulichenko, M. *et al.* Uncertainty-driven dynamics for active learning of interatomic potentials. *Nature Computational Science* **3**, 230–239 (2023).

[148] Finkbeiner, J., Tovey, S. & Holm, C. Generating minimal training sets for machine learned potentials. *Physical Review Letters* **132**, 167301 (2024).

[149] Achar, S. K. *et al.* Reactive active learning: An efficient approach for training machine learning interatomic potentials for reacting systems. *Journal of Chemical Theory and Computation* **21**, 8889–8906 (2025).

[150] Zhang, S. *et al.* Exploring the frontiers of condensed-phase chemistry with a general reactive machine learning potential. *Nature Chemistry* **16**, 727–734 (2024).

[151] Perego, S. & Bonati, L. Data efficient machine learning potentials for modeling catalytic reactivity via active learning and enhanced sampling. *npj Computational Materials* **10**, 291 (2024).

[152] Shuang, F. et al. Modeling extensive defects in metals through classical potential-guided sampling and automated configuration reconstruction. *npj Computational Materials* **11**, 118 (2025). https://doi.org/10.1038/s41524-025-01599-1

[153] Schienbein, P. & Blumberger, J. Nanosecond solvation dynamics of the hematite/liquid water interface at hybrid DFT accuracy using committee neural network potentials. *Physical Chemistry Chemical Physics* **24**, 15365–15375 (2022).

[154] Zeng, J., Cao, L., Xu, M., Zhu, T. & Zhang, J. Z. H. Complex reaction processes in combustion unraveled by neural network-based molecular dynamics simulation. *Nature Communications* **11**, 5713 (2020).

[155] Xing, Z. & Jiang, X. Neural network potential-based molecular investigation of pollutant formation of ammonia and ammonia-hydrogen combustion. *Chemical Engineering Journal* **489**, 151492 (2024).

[156] Yoo, P. *et al.* Neural network reactive force field for c, h, n, and o systems. *npj Computational Materials* **7**, 9 (2021).

[157] Shoghi, N. *et al.* From molecules to materials: Pre-training large generalizable models for atomic property prediction. *arXiv preprint arXiv:2310.16802* (2024).

[158] Zhang, D. *et al.* Pretraining of attention-based deep learning potential model for molecular simulation. *npj Computational Materials* **10**, 94 (2024).

[159] Zhang, D. *et al.* DPA-2: A large atomic model as a multi-task learner. *npj Computational Materials* **10**, 293 (2024).

[160] Zhang, D. et al. A graph neural network for the era of large atomistic models. *npj Computational Materials* **12**, 276 (2026). https://doi.org/10.1038/s41524-026-02146-2

[161] Ramakrishnan, R., Dral, P. O., Rupp, M. & Lilienfeld, O. A. von. Big data meets quantum chemistry approximations: The *Δ*-machine learning approach. *Journal of Chemical Theory and Computation* **11**, 2087–2096 (2015).

[162] Bowman, J. M. *et al.* *Δ*-machine learned potential energy surfaces and force fields. *Journal of Chemical Theory and Computation* **19**, 1–17 (2023).

[163] Kim, J. et al. Data-Efficient Multifidelity Training for High-Fidelity Machine Learning Interatomic Potentials. *Journal of the American Chemical Society* **147**, 1042–1054 (2025). https://doi.org/10.1021/jacs.4c14455

[164] Batatia, I. *et al.* Cross learning between electronic structure theories for unifying molecular, surface, and inorganic crystal foundation force fields. *arXiv preprint arXiv:2510.25380* https://doi.org/10.48550/arXiv.2510.25380 (2025) doi:10.48550/arXiv.2510.25380.

[165] Wood, B. M. *et al.* UMA: A family of universal models for atoms. *arXiv preprint arXiv:2506.23971* (2025).

[166] Ko, T. W. & Ong, S. P. Data-efficient construction of high-fidelity graph deep learning interatomic potentials. *npj Computational Materials* **11**, 65 (2025).

[167] Taniguchi, T. Knowledge distillation of neural network potential for molecular crystals. *Faraday Discussions* **256**, 139–155 (2025).

[168] Messerly, M. *et al.* Multi-fidelity learning for interatomic potentials: Low-level forces and high-level energies are all you need. *Machine Learning: Science and Technology* **6**, 035066 (2025).

[169] Moore, J. H., Cole, D. J. & Csányi, G. Computing Solvation Free Energies of Small Molecules with Experimental Accuracy. *Journal of the American Chemical Society* **148**, 4928–4937 (2026). https://doi.org/10.1021/jacs.5c10940

[170] Kovács, D. P. et al. MACE-OFF: Short-Range Transferable Machine Learning Force Fields for Organic Molecules. *Journal of the American Chemical Society* **147**, 17598–17611 (2025). https://doi.org/10.1021/jacs.4c07099

[171] Levine, D. S., Shuaibi, M., Spotte-Smith, E. W. C., Taylor, M. G., *et al.* The open molecules 2025 (OMol25) dataset, evaluations, and models. *arXiv preprint arXiv:2505.08762* (2025).

[172] Marks, J., Vandezande, J. & Gomes, J. Reliable and Efficient Automated Transition-State Searches with Machine-Learned Interatomic Potentials. arXiv preprint arXiv:2604.00405 (2026). https://doi.org/10.48550/arXiv.2604.00405

[173] Tawfik, S. A. et al. Scale matters: simulation of nanoscopic dendrite initiation in lithium solid electrolyte interphases using a machine learning potential. *Journal of Materials Chemistry A* **13**, 6357–6363 (2025). https://doi.org/10.1039/d4ta08189g

[174] Eckhoff, M. & Reiher, M. Lifelong machine learning potentials. *Journal of Chemical Theory and Computation* **19**, 3509–3525 (2023).

[175] Lim, Y., Park, H., Walsh, A. & Kim, J. Accelerating CO2 direct air capture screening for metal-organic frameworks with a transferable machine learning force field. *Matter* **8**, 102203 (2025).

[176] Kim, J. *et al.* An efficient forgetting-aware fine-tuning framework for pretrained universal machine-learning interatomic potentials. *arXiv preprint arXiv:2506.15223* (2025).

[177] Chen, X. et al. Atomistic learning in the electronically grand-canonical ensemble. *npj Computational Materials* **9**, 73 (2023). https://doi.org/10.1038/s41524-023-01007-6

[178] Bianchi, M. G. *et al.* Electrochemical interfaces at constant potential: Data-efficient transfer learning for machine-learning-based molecular dynamics. *arXiv preprint arXiv:2511.19338* https://doi.org/10.48550/arXiv.2511.19338 (2025) doi:10.48550/arXiv.2511.19338.

[179] Xu, J., Xie, W., Han, Y. & Hu, P. Atomistic insights into the oxidation of flat and stepped platinum surfaces using large-scale machine learning potential-based grand-canonical monte carlo. *ACS Catalysis* **12**, 14812–14824 (2022).

[180] Jung, J. *et al.* Electrochemical degradation of Pt3Co nanoparticles investigated by off-lattice kinetic monte carlo simulations with machine-learned potentials. *ACS Catalysis* **13**, 16078–16087 (2023).

[181] Zhang, Y. & Jiang, B. Universal machine learning for the response of atomistic systems to external fields. *Nature Communications* **14**, 6424 (2023).

[182] Falletta, S. *et al.* Unified differentiable learning of electric response. *Nature Communications* **16**, 4031 (2025).

[183] Wang, R., Fang, S., Huang, Q. & Liu, Y. Constant-Potential Machine Learning Force Field for the Electrochemical Interface. *Journal of Chemical Theory and Computation* **21**, 7628–7635 (2025). https://doi.org/10.1021/acs.jctc.5c00784

[184] Hisama, K., Valadez Huerta, G. & Koyama, M. Molecular dynamics of electric-field driven ionic systems using a universal neural-network potential. *Computational Materials Science* **218**, 111955 (2023).

[185] Ren, F. *et al.* Visualizing the SEI formation between lithium metal and solid-state electrolyte. *Energy & Environmental Science* **17**, 2743–2752 (2024).

[186] Fu, F. *et al.* What drives the heterogeneous interdiffusion in the li-si interfacial region of si anodes: The li flux or the si flux? *npj Computational Materials* **10**, 131 (2024).

[187] Yang, X., Bhowmik, A., Vegge, T. & Hansen, H. A. Neural network potentials for accelerated metadynamics of oxygen reduction kinetics at au-water interfaces. *Chemical Science* **14**, 3913–3922 (2023).

[188] Raman, A. S. & Selloni, A. Acid-base chemistry of a model IrO2 catalytic interface. *The Journal of Physical Chemistry Letters* **14**, 7787–7794 (2023).

[189] Nakanishi, A., Kasamatsu, S., Haruyama, J. & Sugino, O. Defect segregation, water layering, and proton transfer at zirconium oxynitride/water interface examined using neural network potential. *The Journal of Physical Chemistry C* **129**, 2403–2420 (2025).

[190] Cusentino, M., Wood, M. & Thompson, A. Suppression of helium bubble nucleation in beryllium exposed tungsten surfaces. *Nuclear Fusion* **60**, 126018 (2020).

[191] Sumaria, V., Nguyen, L., Tao, F. F. & Sautet, P. Atomic-scale mechanism of platinum catalyst restructuring under a pressure of reactant gas. *Journal of the American Chemical Society* **145**, 392–401 (2023).

[192] Kostiuchenko, T., Körmann, F., Neugebauer, J. & Shapeev, A. Impact of lattice relaxations on phase transitions in a high-entropy alloy studied by machine-learning potentials. *npj Computational Materials* **5**, 55 (2019).

[193] Liu, P., Verdi, C., Karsai, F. & Kresse, G. Phase transitions of zirconia: Machine-learned force fields beyond density functional theory. *Physical Review B* **105**, L060102 (2022).

[194] Li, Z. et al. Navigating polymorph generation and distilled-potential development via entropy-symmetry landscapes for metal plasticity mechanisms. *Nature Communications* **17**, 5070 (2026). https://doi.org/10.1038/s41467-026-73188-9

[195] Allera, A. et al. Activation entropy of dislocation glide in body-centered cubic metals from atomistic simulations. *Nature Communications* **16**, 8367 (2025). https://doi.org/10.1038/s41467-025-62390-w

[196] Bao, Y. et al. Atomistic understanding of hydrogen bubble-induced embrittlement in tungsten enabled by machine learning molecular dynamics. *npj Computational Materials* **12**, 108 (2026). https://doi.org/10.1038/s41524-026-01986-2

[197] Wang, W. et al. Ductility mechanisms in complex concentrated refractory alloys from atomistic fracture simulations. *npj Computational Materials* **11**, 330 (2025). https://doi.org/10.1038/s41524-025-01810-3

[198] Han, K., Deng, B., Barati Farimani, A. & Ceder, G. DistMLIP: A distributed inference platform for machine learning interatomic potentials. *arXiv preprint arXiv:2506.02023* https://doi.org/10.48550/arXiv.2506.02023 (2025) doi:10.48550/arXiv.2506.02023.

[199] Xia, J. & Jiang, B. Efficient parallelization of message passing neural network potentials for large-scale molecular dynamics. *JACS Au* **6**, 3239–3248 (2026).

[200] Fuchs, P., Chen, W., Thaler, S. & Zavadlav, J. Chemtrain-deploy: A parallel and scalable framework for machine learning potentials in million-atom MD simulations. *Journal of Chemical Theory and Computation* **21**, 7550–7560 (2025).

[201] Bochkarev, A., Lysogorskiy, Y. & Drautz, R. Graph atomic cluster expansion for semilocal interactions beyond equivariant message passing. *Physical Review X* **14**, 021036 (2024).

[202] Lysogorskiy, Y., Bochkarev, A. & Drautz, R. Graph atomic cluster expansion for foundational machine learning interatomic potentials. *npj Computational Materials* **12**, 114 (2026). https://doi.org/10.1038/s41524-026-01979-1

[203] Grisafi, A. & Ceriotti, M. Incorporating long-range physics in atomic-scale machine learning. *The Journal of Chemical Physics* **151**, 204105 (2019).

[204] Kosmala, A., Gasteiger, J., Gao, N. & Günnemann, S. Ewald-based long-range message passing for molecular graphs. in *Proceedings of the 40th international conference on machine learning* 17544–17563 (2023).

[205] Wang, Y. *et al.* Neural P3M: A long-range interaction modeling enhancer for geometric GNNs. in *Advances in neural information processing systems* vol. 37 120336–120365 (2024).

[206] Ji, Y., Liang, J. & Xu, Z. Machine-learning interatomic potentials for long-range systems. *Physical Review Letters* **135**, 178001 (2025).

[207] Fuchs, P., Sanocki, M. & Zavadlav, J. Learning non-local molecular interactions via equivariant local representations and charge equilibration. *npj Computational Materials* **11**, 287 (2025).

[208] Zhdanov, M. & Kurenkov, V. Electrostatics from laplacian eigenbasis for neural network interatomic potentials. *arXiv preprint arXiv:2505.14606* (2025).

[209] Bochkarev, A., Lysogorskiy, Y., Ortner, C., Csányi, G. & Drautz, R. Multilayer atomic cluster expansion for semilocal interactions. *Physical Review Research* **4**, L042019 (2022).

[210] Maruf, M. U., Kim, S. & Ahmad, Z. Learning long-range interactions in equivariant machine learning interatomic potentials via electronic degrees of freedom. *The Journal of Physical Chemistry Letters* **16**, 9078–9087 (2025).

[211] Christensen, A. S. & Von Lilienfeld, O. A. On the role of gradients for machine learning of molecular energies and forces. *Machine Learning: Science and Technology* **1**, 045018 (2020).

[212] Zhang, L., Csányi, G., van der Giessen, E. & Maresca, F. Efficiency, accuracy, and transferability of machine learning potentials: Application to dislocations and cracks in iron. *Acta Materialia* **270**, 119788 (2024). https://doi.org/10.1016/j.actamat.2024.119788

[213] NVIDIA Corporation. NVIDIA cuEquivariance GitHub repository. (2024).

[214] Lu, D. *et al.* DP compress: A model compression scheme for generating efficient deep potential models. *Journal of Chemical Theory and Computation* **18**, 5559–5567 (2022).

[215] Zhou, Y. *et al.* FastCHGNet: Training one universal interatomic potential to 1.5 h with 32 GPUs. in *IEEE international parallel and distributed processing symposium (IPDPS)* 236–246 (2025).

[216] Lee, S. Y. *et al.* FlashTP: Fused, sparsity-aware tensor product for machine learning interatomic potentials. in *Proceedings of the 42nd international conference on machine learning* 33143–33156 (2025).

[217] Tan, C. W. *et al.* High-performance training and inference for deep equivariant interatomic potentials. *arXiv preprint arXiv:2504.16068* (2025).